\documentclass[12pt]{article}

\usepackage{amsmath}
\usepackage{amsfonts}
\usepackage{color}
\usepackage{makeidx}       
\usepackage{epsfig}
\usepackage{graphicx}    
\usepackage{hyperref}
\usepackage{caption}
\usepackage{subcaption}
\usepackage{epstopdf}
\usepackage{natbib}
\usepackage{booktabs}

\usepackage[margin=2.0cm]{geometry}

\makeindex    

\title{\begin{flushleft}\textit{Leveraging the network}: a stress-test framework based on DebtRank \end{flushleft}}

\date{ }

\begin{document}

\newgeometry{left=2cm,right=2cm,bottom=2cm,top=2cm}
\thispagestyle{empty}
\vspace{-2cm}

\maketitle

\vspace{-2cm}

%
%
\begin{flushleft}
\indent
 Stefano Battiston,\textsuperscript{a} Guido Caldarelli,\textsuperscript{b} Marco D'Errico,\textsuperscript{a,}\footnote{Corresponding author: \url{marco.derrico@uzh.ch}} Stefano Gurciullo\textsuperscript{c}\\ 

\smallskip

\textsuperscript{a} Department of Banking and Finance, University of Zurich\\
\textsuperscript{b} IMT Alti Studi Lucca, ISC-CNR, Rome, LIMS London\\
\textsuperscript{c} School of Public Policy, University College London \\

\bigskip
\noindent
This version: 
\noindent
\date{\today}

\end{flushleft}

\noindent

\bigskip

\maketitle

\begin{abstract}
We develop a novel stress-test framework to monitor systemic risk in financial systems. The modular structure of the framework allows to accommodate for a variety of shock scenarios, methods to estimate interbank exposures and mechanisms of distress propagation. The main features are as follows. First, the framework allows to estimate and disentangle not only \textit{first-round effects} (i.e. shock on external assets) and \textit{second-round effects} (i.e. distress induced in the interbank network), but also \textit{third-round effects} induced by possible fire sales. Second, it allows to monitor at the same time the \textit{impact} of shocks on individual or groups of financial institutions as well as their  \textit{vulnerability} to shocks on counterparties or certain asset classes. Third, it includes estimates for loss distributions, thus combining network effects with familiar risk measures such as VaR and CVaR.  Fourth, in order to perform robustness analyses and cope with incomplete data, the framework features a module for the generation of sets of networks of interbank exposures that are coherent with the total lending and borrowing of each bank. As an illustration, we carry out a stress--test exercise on a dataset of listed European banks over the years $2008$-$2013$. We find that second-round and third-round effects dominate first-round effects, therefore suggesting that most current stress-test frameworks might lead to a severe underestimation of systemic risk.

\end{abstract}

\renewcommand{\abstractname}{Acknowledgements}
\begin{abstract}
We are grateful to an anonymous referee, and to Joseph Stiglitz, Gabriele Visentin, Serafin Mart\'inez Jaramillo and Irena Vodenska for their useful comments and suggestions on the paper. We also thank the participants in internal seminars at the European Central Bank in Frankfurt (March 2015), the European Systemic Risk Board Joint ATC-ASC Expert Group Meeting on Interconnectedness (May 2015) and NetSci 2015 in Zaragoza. SB and MD acknowledge support from: FET Project SIMPOL nr. 610704, FET project DOLFINS nr 640772, the EU project RMAC nr. 249415, and the Swiss National Fund Professorship grant no. PP00P1-144689. 
\end{abstract}

\section{Introduction}
\label{sec:intro}

The financial crisis has boosted the development of several network-based methodologies to monitor systemic risk in the financial system \citep{eisenberg2001systemic,elsinger2006risk,nier2007network,halaj2013assessing,miranda2013contagion,martinez2012empirical, markose2012too-interconnected,montagna2014contagion,battiston2012debtrank}. 

A traditional approach towards the quantification of systemic risk is to measure the effects of a shock on the external assets of each institution and then to aggregate the losses. However, the crisis has  highlighted that stress-testing should also incorporate so called ``second-round'' effect, which might arise via interbank exposures, either as losses on the asset side or liquidity shortages (see e.g. \cite{bcbs2013liquidity} and references therein).  For instance, the recent ECB comprehensive assessment carried out in 2014 \citep{ecb2014assessment} goes into this direction by taking into account counterparty credit risk while the Basel III framework \citep{basel2010basel} gives  attention to interconnectedness as a key source of systemic risk.

Some network-based methods focus on the events of a bank's default (i.e. its equity going to zero) as the only relevant trigger for the contagion to be passed on to the counterparties. In other words, an institution that has faced some shocks will not affect its counterparties in any way as long as it is left with some positive equity. This is a useful simplification which has allowed for a number of mathematical developments \citep{hurd2011framework}. Because regulators recommend banks to keep their largest single exposure well below their level of equity, most stress test conducted in this way yield essentially to the result that a single initial bank default never triggers any other default. Systemic risk emerges only if, at the same time, one assumes a scenario of weak balance sheets \citep{martinez2012empirical} or a scenario of fire sales \citep{roukny2013default}.

In contrast, both the intuition and the classic Merton approach, suggest that the loss of equity of an institution, even with no default, will imply a decrease in the market value of its obligations to other institutions. In turn, this means a loss of equity for those institutions, as long as they revalue their equity as the difference between assets and liabilities. Therefore, financial distress,  meant as loss of equity, can spread from a bank to another although no default occurs in between.  The total loss of equity in the system can be substantial even if no bank ever defaults in the process. Indeed, in the 2007/2008 crisis, losses due to the mark-to-market re-evaluation of counterparty risk  were much higher than  losses due from  direct defaults.\footnote{See, as a reference, the Basel Committee on Banking Supervision, observing that  “roughly two-thirds of losses attributed to coun- terparty credit risk were due to CVA losses and only about one-third were due to actual defaults.”. \url{http://www.bis.org/ press/p110601.htm}} The so-called DebtRank methodology has been developed with the very idea to capture such a distress propagation \citep{battiston2012debtrank}. The impact of a shock, as measured by DebtRank, is fully comparable to the traditional default-only propagation mechanisms \citep{eisenberg2001systemic,rogers2013failure} in the sense that the latter is a lower bound for the former. In other words, DebtRank measures at least the impact that one would have with the defaults-only, but it is typically larger and this allows to assign a level of systemic importance in most situations in which the traditional method would be unable to do so because the impact would be zero for all banks. 
DebtRank has been applied to several empirical contexts \citep{battiston2012debtrank,diiasio2013capital,tabak2013assessing,poledna2014elimination,fink2014price,aoyama2013debtrank,puliga2014credit} but it was not so far been embedded into a stress-test framework.  In this paper, building on the method introduced in \citep{battiston2012debtrank}, we develop a stress-test \textit{framework} aimed at providing central bankers and practitioners with a monitoring tool of the network effects. The main contributions of our works are as follows. 

First, the framework delivers not only an estimation of \textit{first-round} (shock on external assets), and \textit{second-round} (distress induced in the interbank network) effects, but also a \textit{third-round} effect consisting in possible further losses induced by fire sales. To this end we incorporate a simple mechanism by which banks determine the necessary sales of the asset that was shocked in order to recover their previous leverage level and assuming a linear market impact of the sale on the price of the asset. The three effects are disentangled and can be tracked separately to assess their relative magnitude according to a variety of scenarios on the initial shock on external assets and on liquidity of the asset market. 
Second, the framework allows to monitor at the same time the \textit{impact} and the \textit{vulnerability} of financial institutions. In other words, institutions whose default would cause a large loss to the system become problematic only if they are exposed to large losses when their counterparties or their assets get shocked. These quantities are computed through two \textit{networks of leverage} that are the main linkage between the notion of capital requirements and the notion of interconnectedness. 
Third, the framework allows to estimate loss distributions both at the individual bank level and at the global level, allowing for the computation of individual and global VaR and CVaR (Table \ref{tab:variables}).
Fourth, since data on bilateral exposures are seldom available, the framework includes a module to estimate the interbank network of bilateral exposures given the information on the total lending and borrowing of each bank. Here, we use a combination of fitness model \citep{demasi2006fitness,musmeci2012bootstrapping,montagna2014contagion}, for the network structure and an iterative fitting method to estimate the lending volumes, but alternative methods could be used or added as benchmark comparison (e.g. the maximum entropy method \citep{upper2004estimating,mistrulli2011assessing}, or the minimum density method \citep{anand2014filling}.
Finally, the framework has been developed in MATLAB and is available upon request to the authors. As an illustration, we carry out a stress-test exercise on a dataset of 183 European banks over the years $2008$--$2013$, starting from the estimation of their interbank exposures.

This paper is organized as follows. In Section \ref{sec:related work} we review similar or related work; in Section \ref{sec:stress-test} we describe the main aspects of the framework, providing an outline of the distress process, a discussion of the main variables, and the framework's building blocks; in Section \ref{sec:exercise}, we show how the framework can be applied to a dataset and we discuss the main results of this exercise; in Section \ref{sec:discussion}, we review the main contributions and introduce elements for future research. In Appendix \ref{sec:methods}, we provide the technical details of the distress propagation process, including how the key measures are computed; in the Appendix \ref{sec:data_collection}, we described the data we used for the exercise  in Section \ref{sec:exercise} and, last, in Appendix
\ref{sec:network_reconstruction}, we outline the network reconstruction methods when only the total interbank lending/borrowing for each bank is known.

\subsection{Related work}
\label{sec:related work}
The recent  -- and still ongoing -- economic and financial crisis has made clear the importance of methods of early detection of systemic risk in the financial system. In particular, researchers, regulators and policy-makers have recognized the importance of adopting a macroprudential approach to understand and mitigate financial stability. Notwithstanding the many efforts \citep{kolb2013lessons}, regulators still lack an adequate framework to measure and address systemic risk\footnote{In the following, we refer to \textit{systemic risk} to indicate the probability that a large portion of the financial system is in distress or collapses.}. 

The traditional micro-prudential approach consists in trying and ensuring the stability of the banks, one by one, with the assumption that as long as each unit is safe the system is safe. This approach has demonstrated to be a dangerous over-simplification of the situation \citep{borio2003macroprudential}. Indeed, we have learned that it is precisely the interdependence among institutions, both in terms of liabilities or complex financial instruments and in terms of common exposure to asset classes what leads to the emergence of systemic risk and makes the prediction of the behaviour of financial systems so difficult \citep{battiston2012credit}. While risk diversification at a single institution can indeed lower its individual risk, if all institutions behave in a similar way, herding behaviour can instead amplify the risk. Clearly, if all banks take similar positions, the failure of one bank can cause a global distress \citep{brock2009more,stiglitz2010risk,caccioli2013how}, because of the increased sensitivity to price changes \citep{patzelt2013inherent}. To add more complexity, the causes of market movements are still under debate \citep{cutler1989moves,cornell2013moves}, suggesting that exogenous instabilities add up to endogenous ones \citep{danielsson2012endogenous}. 
The tension between individual regulation and global regulation \citep{beale2011individual} poses a series of challenging questions to researchers, practitioners and regulators \citep{boe2013stresstesting}.

Traditionally, well before the recent crisis, it was argued that systemic risk is real when contagion phenomena across countries take place \citep{krugman1991international,bordo1995real}. In this spirit, a series of studies dealt with the description of systemic risk in the financial system from the perspective of the contagion channels across balance-sheet of several institutions \citep{elsinger2006risk,gai2011complexity,miranda2013contagion, huang2013cascading,montagna2014contagion,glasserman2015likely}. 
In particular, some focus was drawn upon the topology 
of connections (or the network \citep{caldarelli2007scalefree}) between institutions \citep{eisenberg2001systemic,roukny2013default,acemoglu2013systemic}.

In this way, the problem of analysing systemic risk splits in two distinct problems \citep{cont2010network}. First, the problem of understanding the role of an opaque (if not unknown) structure of financial contracts \citep{caldarelli2013reconstructing} and, second, the problem of providing a measure for the assessment of the impact of a given shock \citep{battiston2012debtrank}. As for the first problem, the obvious starting point is to consider the structure of the interbank network
\citep{demasi2006fitness,iori2008network,may2010systemic,mistrulli2011assessing,roukny2014network}, with the aim of possibly extracting some early warning signals \citep{squartini2013early}. While many argued that the network structure can be intrinsically a source of instability, it turns out instead that no specific 
topology can be considered as systematically safer than the others \citep{roukny2013default}. Indeed, only the interplay between market liquidity, capital requirements and network structure can help in the understanding of the systemic risk \citep{roukny2013default, loepfe2013towards}. For the second problem, researchers have tried to describe the dynamics of propagation of defaults with various methods, including by means of agent-based models \citep{geanakopios2012getting} or by modelling the evolution of financial distress across balance-sheets conditional upon shocks in one or more institutions \citep{battiston2012debtrank}. 

From the perspective of financial regulations, capital requirements represent the cornerstone of prudential regulations. Institutions are required to hold capital as a buffer to shocks of any nature. The most used risk measures (such as Value at Risk and Expected Shortfall) are indeed related to the quantity of cash each individual bank needs to set aside in order to cover the \textit{direct} exposures to different types of risk. In such manner, the \textit{indirect} exposures arising from the interconnected nature of the financial system are not considered. Interconnectedness, though, is now entering the  debate on regulation: for example, the definition of ``Global Systemically Important Banks''  \citep[G-SIBs,][]{bis2011gsib} does include the concept of interconnectedness, thereby measured as the aggregate value of assets and liabilities each bank has with respect to other banking institutions. Although this represents a fundamental step towards the inclusion of interconnectedness in assessing systemic risk, a further level of disaggregation would be needed. In fact, institutions that are similar in terms of their aggregated exposures (including those \textit{vis-\`a-vis} other financial institutions), might have completely different sets of counterparties, therefore implying different levels of systemic impact and/or vulnerability to shocks. Another important point is that the potential negative effects arising from interconnectedness ought to be included into the definition of capital requirements.

\section{The DebtRank stress-test framework}
\label{sec:stress-test}

In this Section, we introduce and  describe the DebtRank stress-test framework. One of the main characteristics of the framework lies in its \textit{flexibility} along the following four main dimensions.

\begin{enumerate}
\item \textbf{Shock type}. The framework can implement different shock types and scenarios (on external assets).
\item \textbf{Network estimation}. When detailed bilateral interbank exposures are not available, the framework provides a module to estimate the interbank network from the total interbank assets and liabilities of each bank, 
\item \textbf{Contagion dynamics}. The framework can implement two different contagion dynamics,  distress contagion and default contagion.
\item \textbf{Systemic risk indicators}. The framework returns as output a series of systemic risk indicators, both at the individual and a the global level. The user can aptly combine this information to extract the information needed. Several graphical outputs are also available and represent a key feature of the framework: graphics are specifically designed to capture relevant information at a glance.
\end{enumerate}

Given the flexibility of the framework and the number of outputs produced, in the remainder of the Section, we focus on:
\begin{enumerate}
\item describing the main features of the DebtRank \textit{distress process} as the key foundation of the framework;
\item providing a qualitative description of the main variables of interests;
\item providing a technical summary of the \textit{building blocks} of the framework, which include the \textit{inputs} required, the \textit{outputs} that can be obtained and the different \textit{modules} constituting the framework.
\end{enumerate}

The reader can find detailed information about the process and the main variables of interest in the methodological appendix \ref{sec:methods}.

\subsection{Outline of the distress process}

One of the key concerns in the measurement of systemic risk is to quantify \textit{losses} at the individual and global level. In particular, DebtRank focuses on the depletion of equity when banks experience losses in external or interbank assets. We envision a system of $n$ banks (indexed by $i = 1, \ldots, n$) and $m$ external assets (indexed by $k = 1, \ldots, m$). The framework features a dynamic distress model, with $t = 0, 1, \ldots, T, T+1, T+2$:

\begin{table}[h!]
\centering
\caption{The distress dynamics.\label{tab:dynamics}}
\begin{tabular*}{\textwidth}{lll}
Time & Round &  Effects on balance sheets \\ \toprule
$t=0$ & Baseline  & Initial allocation\\ \vspace{-0.3cm} \\
$t=1$ & First round effects &  {\begin{tabular}[x]{@{}l@{}} Shocks on external assets;\\  immediate write-off on balance sheets \end{tabular}}\\ \vspace{-0.3cm} \\
$t=2$ & Second round begins & {\begin{tabular}[x]{@{}l@{}} Reverberation on the interbank  lending network; \\ banks receive the distress of their neighbors \end{tabular} }\\ 
$t = T$ & Second round ends & Second round effects \\
$t = T+1$ & Third round begins & Banks aim at restoring original leverage value\\
$t = T + 2$& Third round ends & Final effects\\
\toprule
\end{tabular*}
\end{table}

\begin{description}
\item[Initial configuration.]  At time $t=0$, banks allocate their uses and sources of funding, all variables at this time represent the initial conditions of the process.
\item[First round.] At time $t=1$, we assume a negative shock on the value of one or more assets $k$. Banks immediately record the loss and, as they have to pay back their liabilities, reduce their equity level accordingly. We refer to these losses in equity as \textit{first round} effects.
\item[Second round.] Given the equity loss of each bank, the likelihood of a bank repaying its obligations on the interbank lending market becomes lower, therefore reducing the market value of its obligations. This triggers effects on the interbank lending network. Indeed, from  $t=2$ to $t = T \geq 2$, we model the propagation of distress in the interbank network. We refer to the loss on equity at  this point as \textit{second round} effects. At at certain time  $ t = T$, the second round ends.
\item[Third round.] From time $ t = T + 1$, the equity level is reduced from the initial configuration and banks aim at restoring the original leverage levels. In order to do so, they sell external assets (fire sales). This triggers further effects on the price of external assets and  reduces equity levels to a greater extent. We refer to  these losses as \textit{third round} effects.
\end{description}

Our framework is based on the clear separation between rounds of distress. At each round, the \textit{loss in equity} is the key variable in our framework. As a quick reference, a summary of the distress dynamics is provided in Table \ref{tab:dynamics}.

\subsection{Measuring systemic risk: the main variables}

We now give a brief description of the main variables in the framework, and their interpretation in terms of systemic risk. As a reference, the reader can find a summary of these variables in Table \ref{tab:variables}.

\begin{table}[h!]
\centering
\caption{Description of the main variables in the stress-test framework.\label{tab:variables}}
\begin{tabular*}{\textwidth}{llll}
\textbf{Name} & \textbf{Symbol} & \textbf{Ref.} &\textbf{Explanation} \\ \toprule {\begin{tabular}[x]{@{}l@{}}Individual\\ vulnerability at $t$ \end{tabular}} & $h_i(t)$ & Eq. \ref{eq:individual_equity_loss} & Relative loss in equity of bank $i$ (up to time $t$).\\ \vspace{-0.3cm} \\{\begin{tabular}[x]{@{}l@{}}Global \\ vulnerability at $t$ \end{tabular}}  & $H(t)$ & Eq. \ref{eq:global_equity_loss} & Relative loss on equity for the whole system (up to time $t$).\\ \vspace{-0.3cm} \\
 Individual impact & $DR_i$ & Eq. \ref{eq:impact} &  {\begin{tabular}[x]{@{}l@{}}Total relative loss on equity  \\  \textit{induced} by the default of $i$ on the whole network.\end{tabular}} \\ \vspace{-0.3cm} \\
{\begin{tabular}[x]{@{}l@{}}Individual Value\\ at Risk at $t$ \end{tabular}}  &  $VaR^{\alpha}_{i}(t)$ & Eq. \ref{eq:var} & {\begin{tabular}[x]{@{}l@{}}Value at Risk at level $\alpha$  \\ for the individual loss distribution of institution $i$.\end{tabular}} \\ \vspace{-0.3cm} \\
{\begin{tabular}[x]{@{}l@{}}Global Value\\ at Risk at $t$ \end{tabular}}  & $VaR^{\alpha}_{\text{glob}}(t)$ & Eq. \ref{eq:globcvar} & {\begin{tabular}[x]{@{}l@{}}Value at Risk at level $\alpha$  \\ for the global relative loss distribution on equity.\end{tabular}} \\
\toprule
\end{tabular*}
\end{table}

\paragraph{Vulnerability} As previously noted, the key quantity in the framework is the \textit{loss in equity} for each bank at each time $t$. In terms of systemic risk, however, there is substantial difference between the loss in equity a bank \textit{suffers} and the loss in equity a bank \textit{induces} in the system. We call the first variable the \textit{vulnerability} of a bank and the second variable the \textit{impact} of a bank onto the system as a whole. More formally,  given the equity values at the initial configuration $E_i(0)$, we define the \textit{individual vulnerability} $h_i(t)$ of bank $i$ at $t$ as follows: 

\begin{equation}
\nonumber
\tag{individual vulnerability}
h_i(t) = \min \left\{1, \frac{E_i(0) - E_i(t)}{E_i(0)}\right\}.
\end{equation}

The bank defaults when $h_i(t) = 1$.  Similarly, we can compute the \textit{global vulnerability} of the system at time $t$, by taking the weighted average of $h_i(t)$, with weights  given by the relative initial equity:

\begin{equation}
\nonumber
\tag{global vulnerability}
H(t) = \sum_{i = 1}^n \left( \frac{E_i(0)}{\sum_j E_j(0)} h_i(t) \right).
\end{equation}

\paragraph{Impact.} Institutions in a financial system are not only systemically relevant in terms of the shock they receive but also in terms of the loss they cause in case of their default. We call the \textit{individual impact} of an institution $i$, the relative equity loss \textit{induced} by the default of $i$ (as computed in Equation \ref{eq:impact} in the methodological appendix \ref{sec:methods}). We denote the impact with $DR_i$ as it is consistent with the original DebtRank approach introduced in \citep{battiston2012debtrank}. Notice that the measure of impact naturally applies only to the distress a bank induces in the interbank network. 

\paragraph{Loss distributions.} Conditioning to specific shocks, one can characterize a loss distribution both at the individual $h_i(t)$ and at the global level $H(t)$ at each time $t$. In this context, ``loss'' and ``vulnerability'' can be used interchangeably. Notice that both the notions of individual and global loss distribution are key aspects in the quantification of systemic risk. As a matter of fact, a large fraction of the global losses may be attributable to a few key banking institutions.  In particular, we compute the Value at Risk (VaR) and the Conditional Value at Risk (CVaR), as these measures have emerged as some of the key tools for risk assessment. In our framework, these measures move towards the inclusion of network effects. In addition, the global loss distribution provides a clear understanding of the vulnerability of the system as a whole conditional to a specific shock.

\paragraph{Evolution in time.} All measures of vulnerability/losses and impact both at the individual and global level can be tracked over time, therefore providing a way to monitor the evolution of key figures in terms of systemic risk. In the exercise reported in Section \ref{sec:exercise}, we focus on the monitoring of these key variables for a subset of $183$ European banks in the years from $2008$ to $2013$. The dynamics of these key systemic risk variables allows to capture the evolution of systemic risk in time.

\subsection{The framework's building blocks}

Since the DebtRank stress-test framework features several quantitative and graphical outputs for input data that are usually publicly available, we now provide a brief, yet comprehensive, overview of the main building blocks. We use Table \ref{tab:building_blocks} as the main reference. 

\begin{table}[h!]
\caption{Building blocks of the stress-test framework\label{tab:building_blocks}}
\centering
\textbf{Building blocks of the stress-test framework}
\begin{tabular}{lll}
\toprule
& & \\
\textbf{Input} & Banks' balance sheets $\rightarrow$ & \begin{tabular}{|l}
i) lending / borrowing (interbank vs total) \\ 
ii) external assets (with possible breakdowns) \\
iii) equity (and reserve capital in general)
\end{tabular}\\
& & \\
& Shock scenario $\rightarrow$ & \begin{tabular}{|l}
i) one or more banks \\ ii) one or more asset classes
\end{tabular}\\  & & \\ \hline
& & \\
\textbf{Output} & Results of Modelling scenario $\rightarrow$ & \begin{tabular}{|ll}
Contagion & \begin{tabular}{l}
DebtRank \\ Default Cascade
\end{tabular} \\
& \\
Exposure estimation & \begin{tabular}{l}
Fitness model \\
(Null models) (1 \& 2) \\
(Maximum entropy) \\
(Minimum density)
\end{tabular}
\end{tabular}
\\ \toprule
\end{tabular} 	
\end{table}

\subsubsection{Input}

\paragraph{Input - data on balance sheets.} The fundamental input data are represented by banks' balance sheets. In particular, the framework takes the equity, the total asset value and the total interbank lending and borrowing of each bank as minimal inputs. More granular data on the structure of external assets are indeed possible (e.g. in case one wants to simulate a shock on a specific asset class).

\paragraph{Input - Shock scenario.} The flexibility of the modeling framework allows for a number of shock scenarios, including: 
\begin{enumerate}
\item a fixed shock (e.g. $1\%$) on the value of all external assets;
\item a shock on the value of all external assets drawn from a specific probability distribution (e.g. a Beta distribution, which we use in the exercise in Section \ref{sec:exercise}.);
\item when more detailed information on the holdings in external assets for banks is available, the shock (either fixed or drawn from a probability distribution) on specific asset classes.\footnote{This also allows to run the stress-test by applying heterogenous shocks with a pre-determined correlation structure. However, we will tackle this issue more specifically in future works.}
\end{enumerate}

\subsubsection{Output}

\paragraph{Output - results} As outlined above, the framework allows to compute the main systemic risk variable for two main types of contagion dynamics: \begin{enumerate}
\item the \textit{default cascade} dynamics: banks impact other banks only in case of their default (see, for the technical details, the discussion related to Equation \ref{eq:main_debt_rank} in the methodological appendix \ref{sec:methods}.)
\item the \textit{DebtRank} dynamics: banks impact other banks regardless of whether the event of default occurred. The rationale behind this type of dynamics is that, as banks reduce their equity levels to face losses, they decrease their distance to default and therefore are less likely to repay their obligations. In this case, the market value of their obligations is reduced and is hence reflected on the asset side of their counterparties in the interbank market.
\end{enumerate}

\paragraph{Output - bilateral exposures estimation.} As detailed data on banks' bilateral exposures are often not publicly available, estimations need to be performed in order to run the framework. Even though such estimations constitute a key input of the stress test framework in case the exposures are not known,  they constitute an output on their own, because they can be then analyzed with the typical tools of network analysis. Also, the estimations can serve for two other purposes: i) as a benchmark for comparison with the observed data, \textit{\`a la} \citet{savage1960statistical}, or ii) for the estimation of missing data \citep{anand2014filling}. From a technical viewpoint, the methodology we use to estimate the interbank network is based on the so called ``fitness model'' \citep{demasi2006fitness, musmeci2012bootstrapping}. The technical details are reported in Appendix \ref{sec:network_reconstruction}.

\section{The framework at work: results of a stress test exercise\label{sec:exercise}}

In order to show how the framework works and what type of outputs are available, in this Section we apply the framework to a specific dataset of $183$ EU banks for the years $2008-2013$. More details on the dataset are available in Appendix \ref{sec:data_collection}. In brief:

\begin{enumerate}
\item We collect yearly data on equity, external assets, interbank assets and liabilities for the set of banks under scrutiny;
\item We estimated the exposures by combining the fitness model and an interative fitting procedure (Appendix \ref{sec:network_reconstruction}), generating (for each year) $100$ networks compatible with the total interbank borrowing and lending of each bank at end-year;
\item We then ran the stress-test in order to obtain the main systemic risk variables for all years. When not explicitly specified, the statistics reported in this Section are computed by taking the median value of the $100$ networks.
\end{enumerate}
In the remainder of this Section, we describe the main results, including some key charts and figures, in order to show part of the graphical output of the framework.

\subsection{Vulnerability and impact}

\begin{figure}[t]
\centering
\includegraphics[width = 0.49\textwidth, trim = 0cm 0.0cm 0 0]
{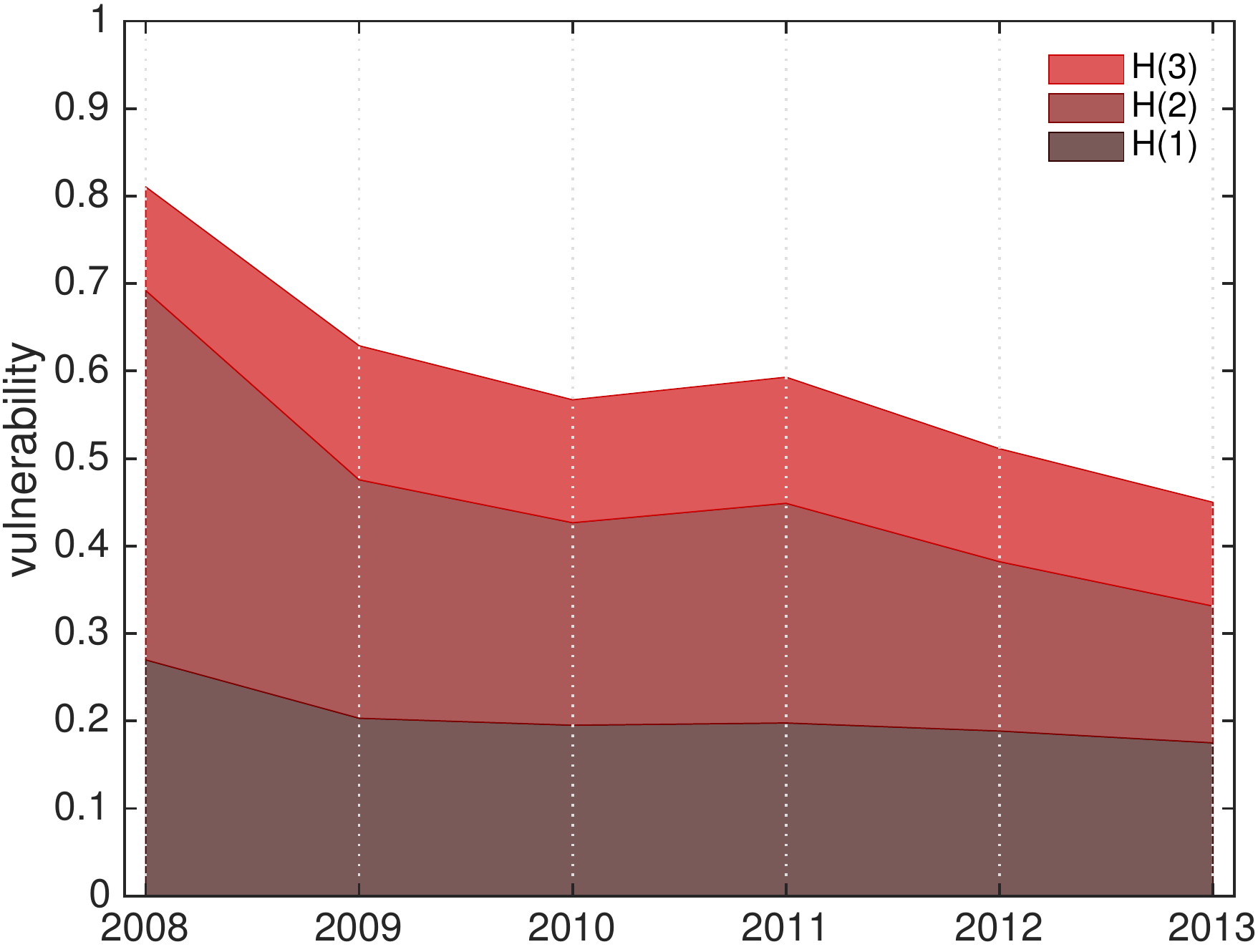}
\includegraphics[width = 0.49\textwidth]{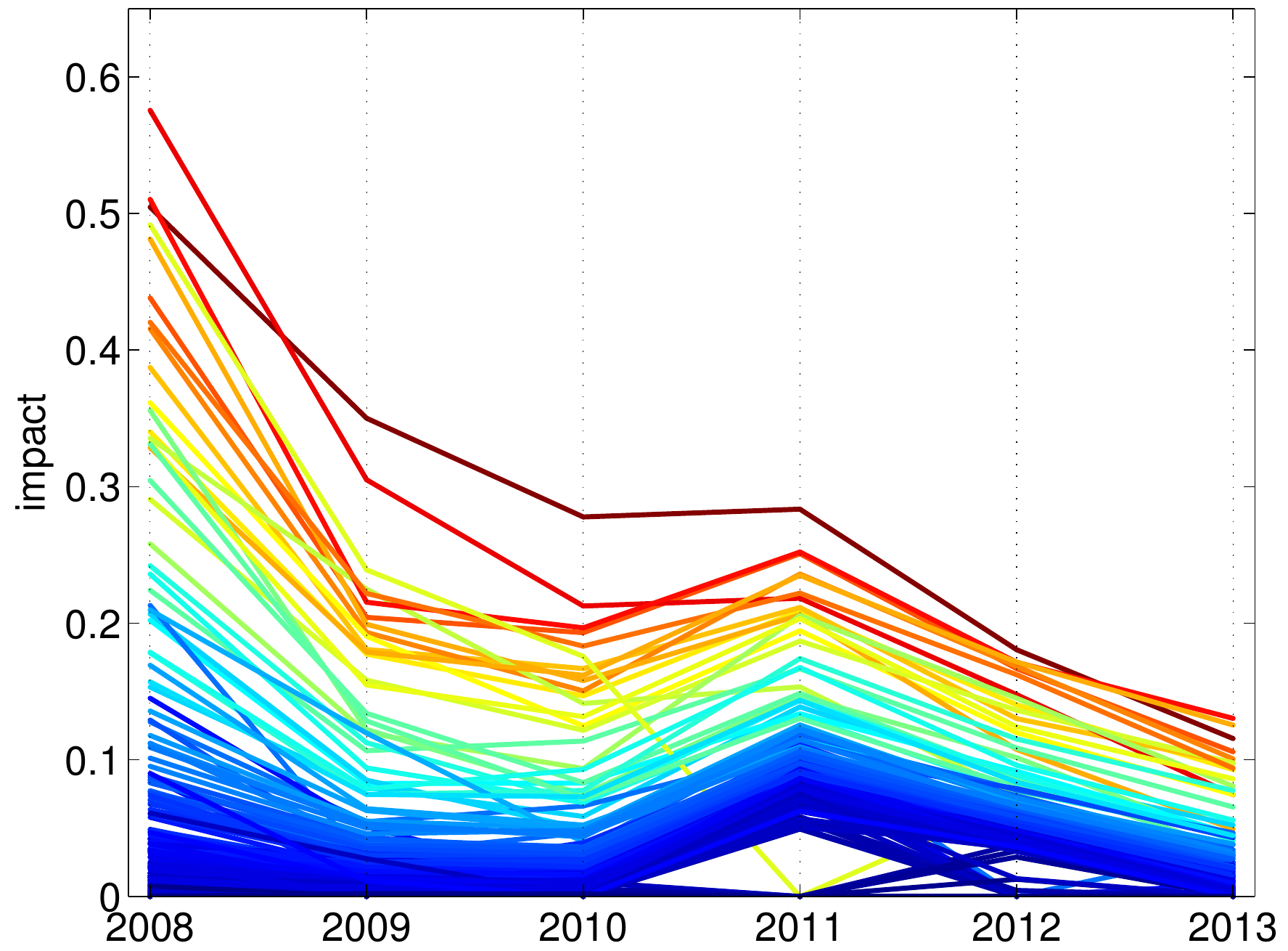}
\caption{Systemic vulnerability and individual impact over time. (Left) Plot of the global vulnerability in time and its decomposition w.r.t. the different rounds. (Right) Individual impact over time. In order to show that impactful institutions keep being so during the years, colors reflect the impact in $2008$.\label{fig:vulner_impact}}
\end{figure}

Figure \ref{fig:vulner_impact} provides an overview of the response of the reconstructed financial networks and its individual elements to the distress scenarios simulated. The chart on the left shows the dynamics of global equity losses ($H$) from $2008$ to $2013$, the values reported are the median value of $H$ across the $100$ networks in the Monte Carlo sample and are computed for a common shock of $1\%$ on the external assets.  The chart also offers a deconstruction of the losses, according to if they are caused by the first (external assets shocks), second (reverberation on the interbank lending network), and third (fire sales) round of distress propagation. The relative losses in equity due to the second and third rounds are substantial, implying that an assessment of systemic risk solely based on first order effects is bound to underestimate potential losses.  The chart on the right shows the evolution of the impact for each of the $183$ banks in the sample throughout the years. Each line is the median of the impact calculated over the $100$ networks in the ensemble.  The plot clearly shows a general decrease in the systemic impact for the individual institutions over time.  In order to visually capture the persistency over time of banks with higher or lower impact, the colours reflect the level of the average impact computed over the years. In particular, red lines are associated to banks that consistently show a high impact. Conversely, blue lines are associated to banks that have a consistently low impact. We observe a certain level of stability of the relative levels: banks which show a higher systemic impact tend to do so throughout the years.

\begin{figure}[h!]
\includegraphics[width = 0.49\textwidth, trim = 0cm 0cm 0.6cm 0]{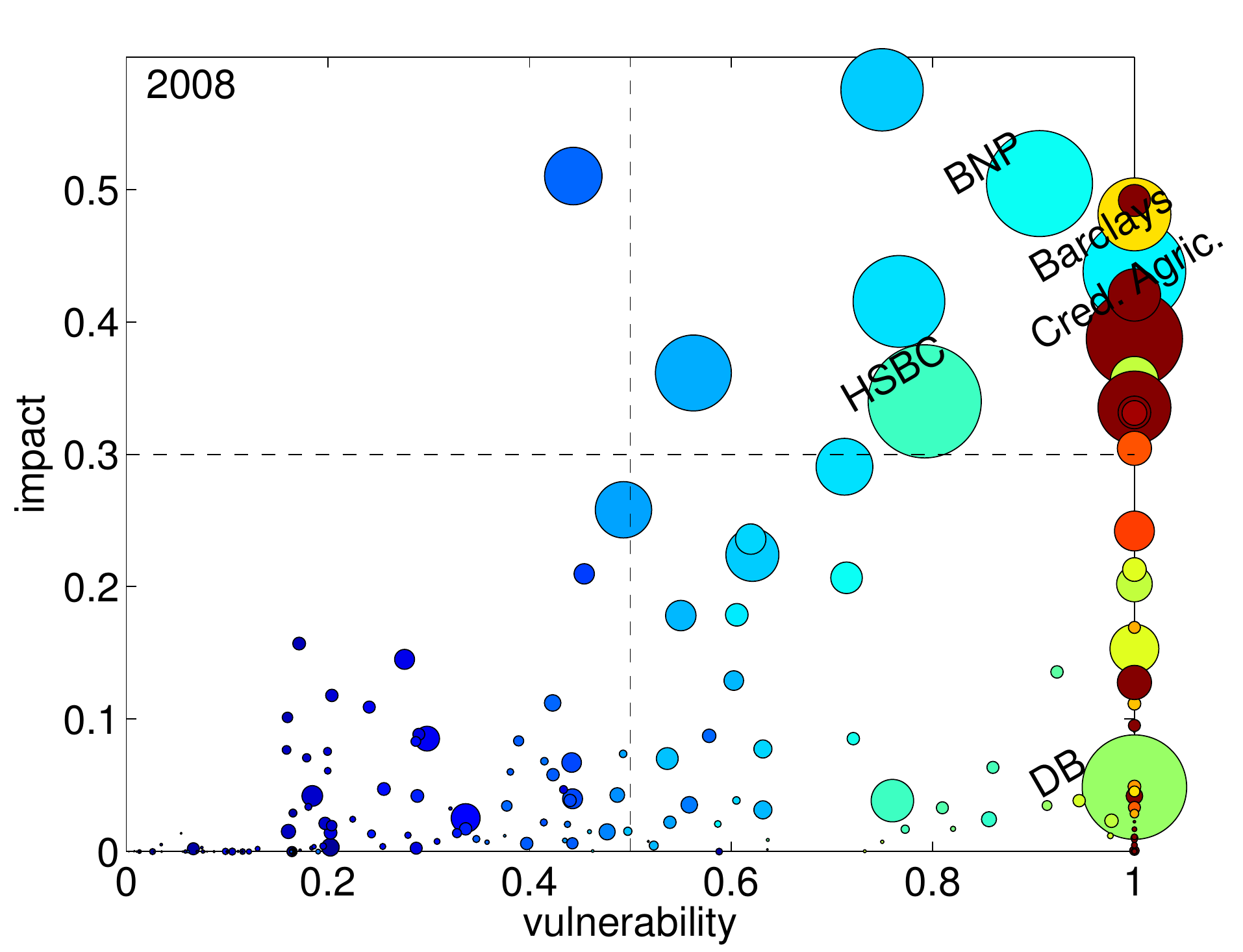}
\includegraphics[width = 0.49\textwidth]{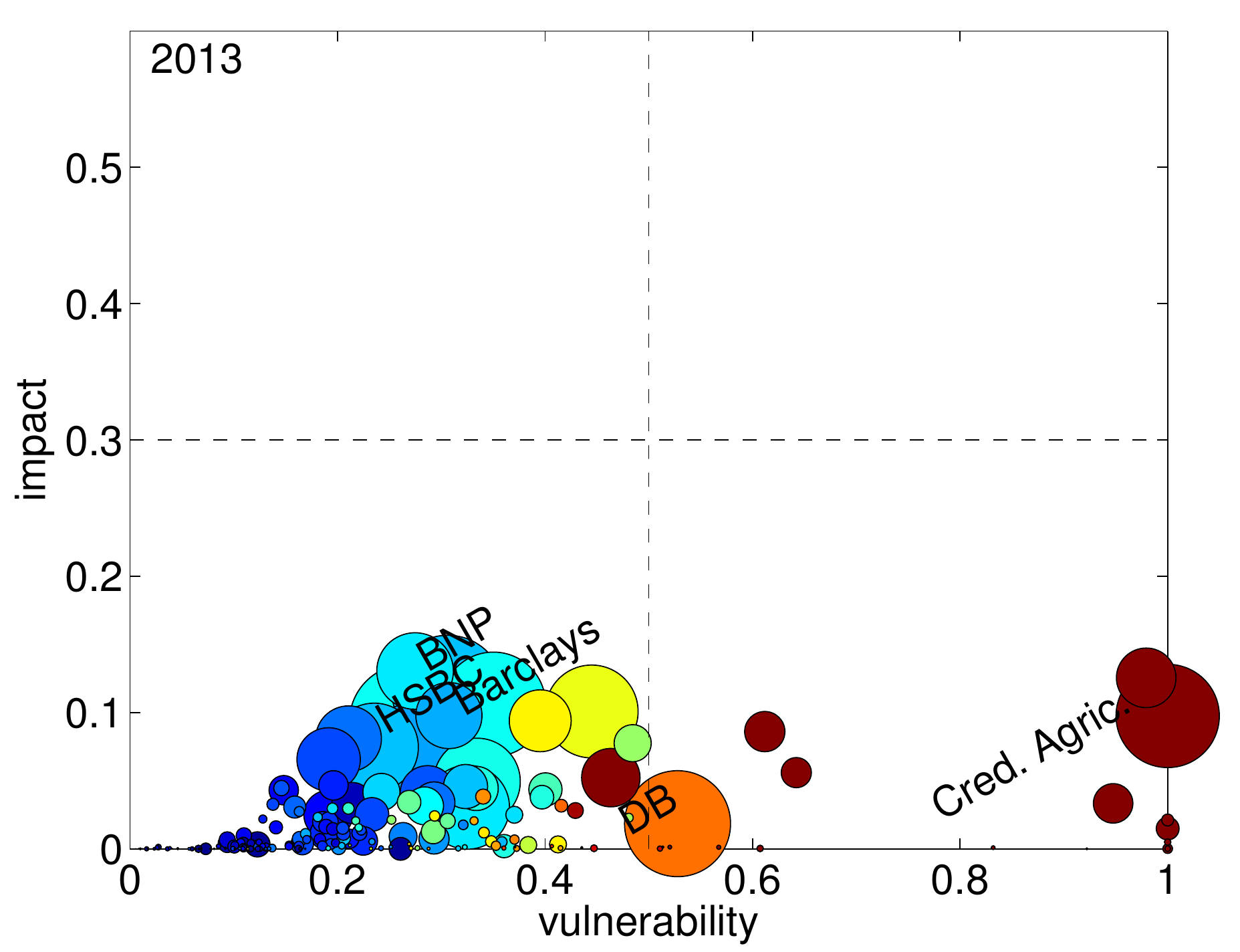}
\caption{Individual vulnerability vs individual impact (2008 and 2013) Circle size reflects asset size, colors reflect the magnitude of the interbank leverage. The four quadrants divide the banks into four categories.\label{fig:vuln_vs_imp}}
\end{figure}

From a systemic risk perspective, it is of particular interest to compare the two main systemic risk quantities associated to each individual bank: the vulnerability to external shocks and the impact of a bank onto the system in case of its default. By \textit{jointly} analyzing these two quantities, we divide institutions into four main categories: i) high vulnerability / high impact,
ii) high vulnerability / low impact, iii) low vulnerability / low impact, iv) low vulnerability / high impact.

Results for this exercise are reported in Figure \ref{fig:vuln_vs_imp}. The graphs report a plot of the vulnerability $h_i$ at the second round versus the impact $DR_i$ for each year in the sample. The $[0, 1] \times [0, 1]$ square is divided into four quadrants, which correspond to the aforementioned four categories. Interbank leverage and total asset size are respectively visualised by node colour (red implies high leverage, blue otherwise) and node size. Both interbank leverage and asset size appear to be associated with high values of vulnerability and impact.
We observe an interesting phenomenon: in $2008$, a high number of large (in terms of asset size) institutions are \textit{both} highly vulnerable (up to their default) and impactful (up to $70\%$ of the total initial equity). Their systemic relevance is therefore extremely high, as they have higher likelihood to receive distress. In turn, once the distress has been received, they would have a great impact on the rest of the system. The situation improves over time and, in $2013$, no bank is in the upper right quadrant.  Some financial institutions retain, though, very high vulnerability and significant impact. A financial institution that can cause a global relative equity loss of 10\% still acts as a source of systemic risk not to be ignored.  However, some large institutions are still prone to receive high level of distress, and nevertheless keep a significant impact (up to $20\%$ on the rest of the system).  We also notice that those institutions which are both vulnerable and impactful are generally large and very large ones in terms of asset size.

 \subsection{Decomposition of $1^{\text{st}}$ and $2^{\text{nd}}$ round effects}
 
\begin{figure}[h!]
\includegraphics[width = 0.49\textwidth]{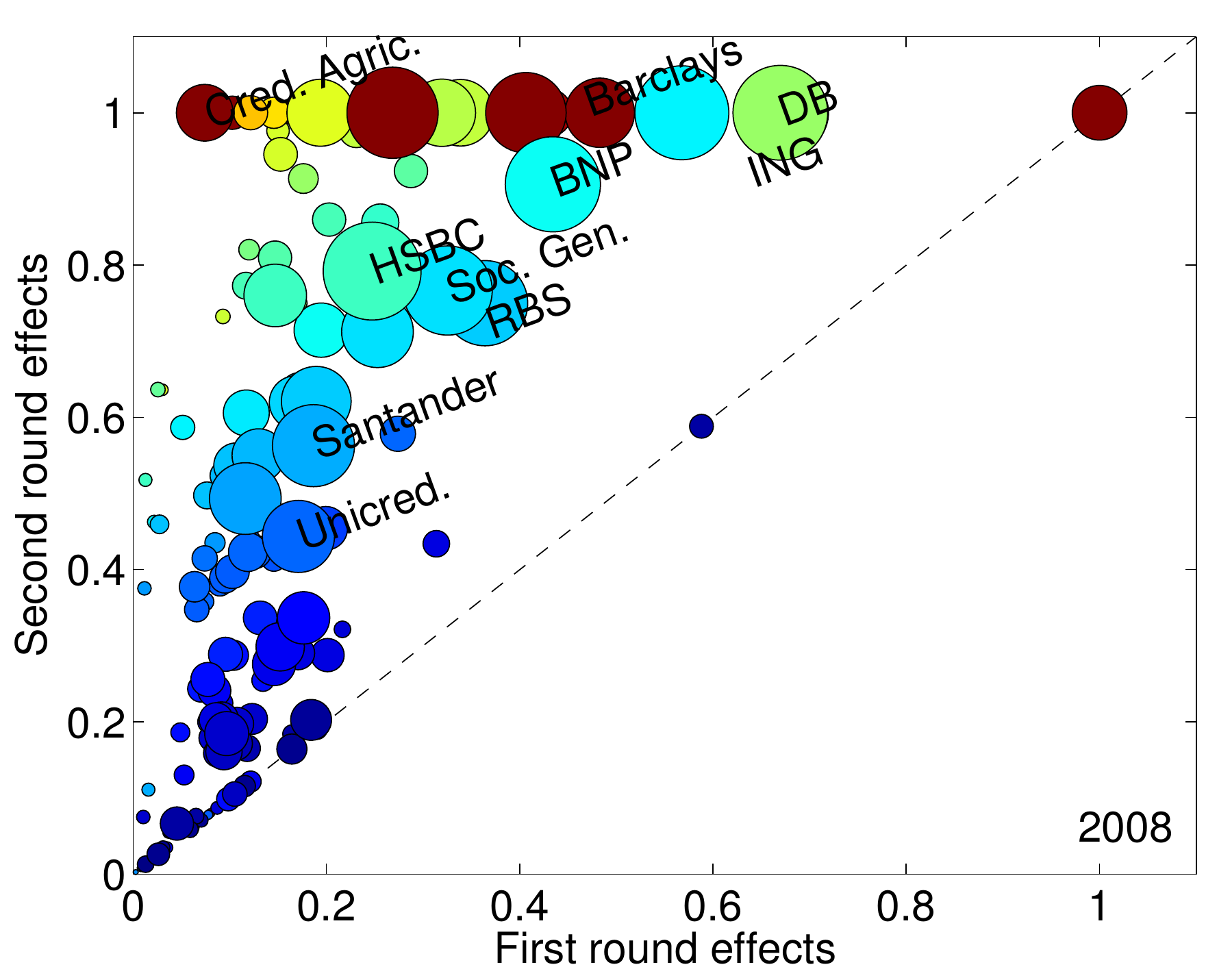}
\includegraphics[width = 0.49\textwidth]{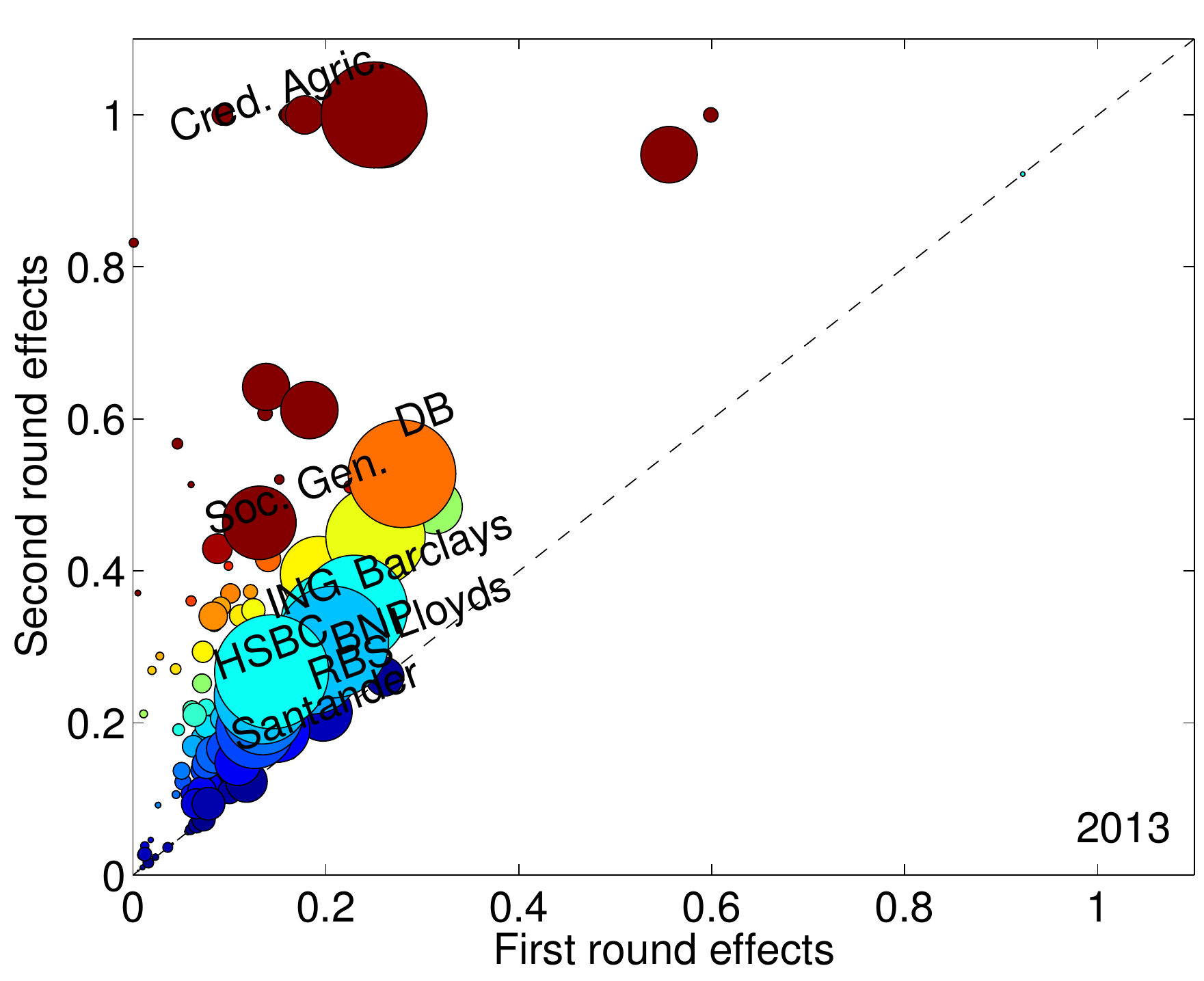}
\caption{Decomposition of first and second round effects in 2008 and 2013 for an initial shock on external assets $r(1) = 0.01$. The names of the first top ten institutions by asset size for each year are shown. \label{fig:eff_decomp}}
\end{figure}

Figure \ref{fig:eff_decomp} shows a way of visualizing the decomposition of first and second round effects. Again, we compare the years 2008 (left) and 2013 (right). The x-axis plots the losses at the first round and y-axis the losses after the second round. Since the losses at the second round \textit{include} the ones at the first, points must lie above the line bisecting the first quadrant. Nodes lying on the line itself are isolated in all the artificially generated networks.  We observe a significant reduction in the effects. As usual, the color reflects the interbank leverage and circle diameter the asset size. Consistently with the findings in Appendix \ref{sec:methods}, nodes with higher interbank leverage typically suffer more losses in the second round.

\subsection{Distribution of losses}
\subsubsection{Global losses}
\begin{figure}
\includegraphics[width = 0.49\textwidth]{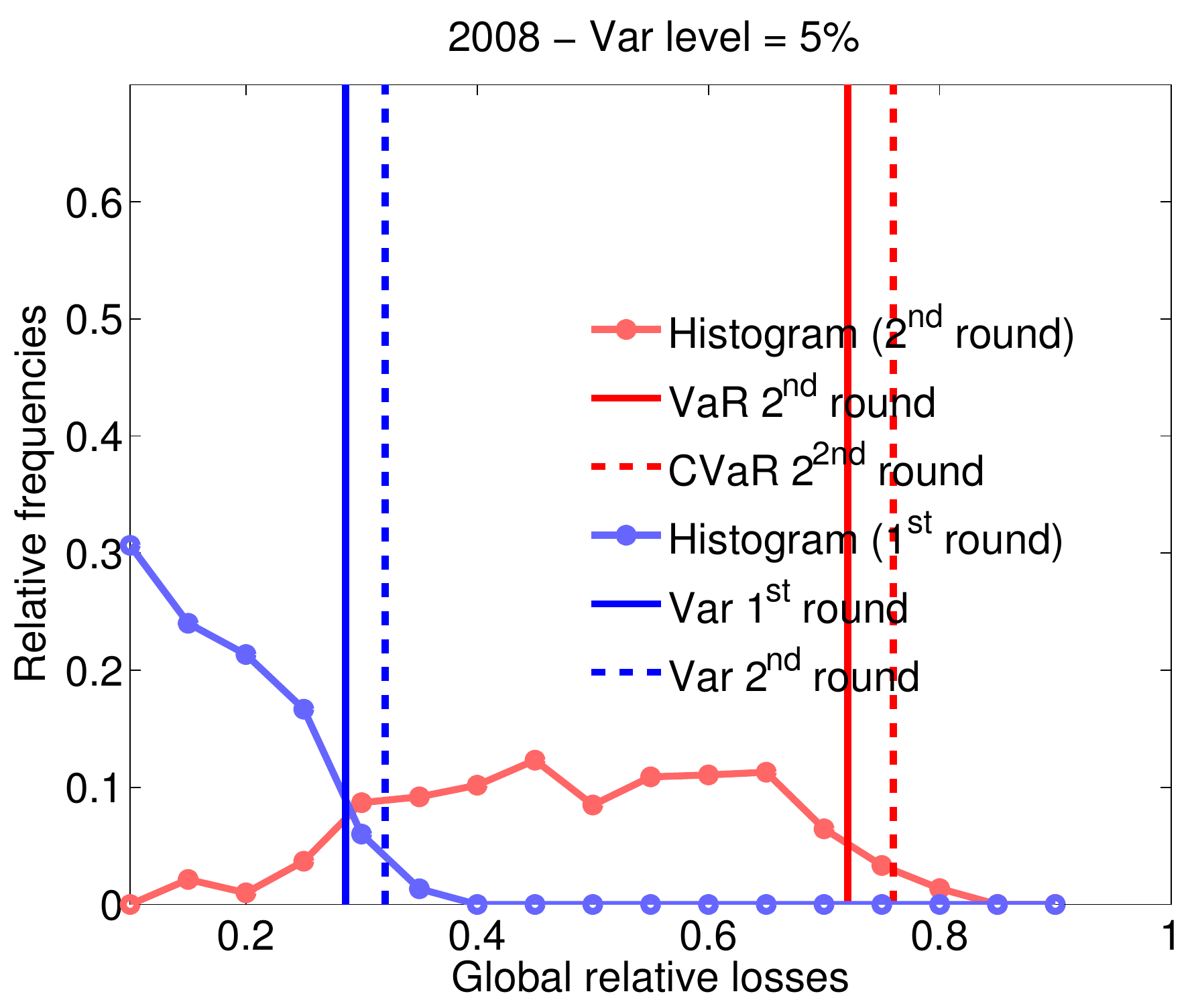}
\includegraphics[width = 0.49\textwidth]{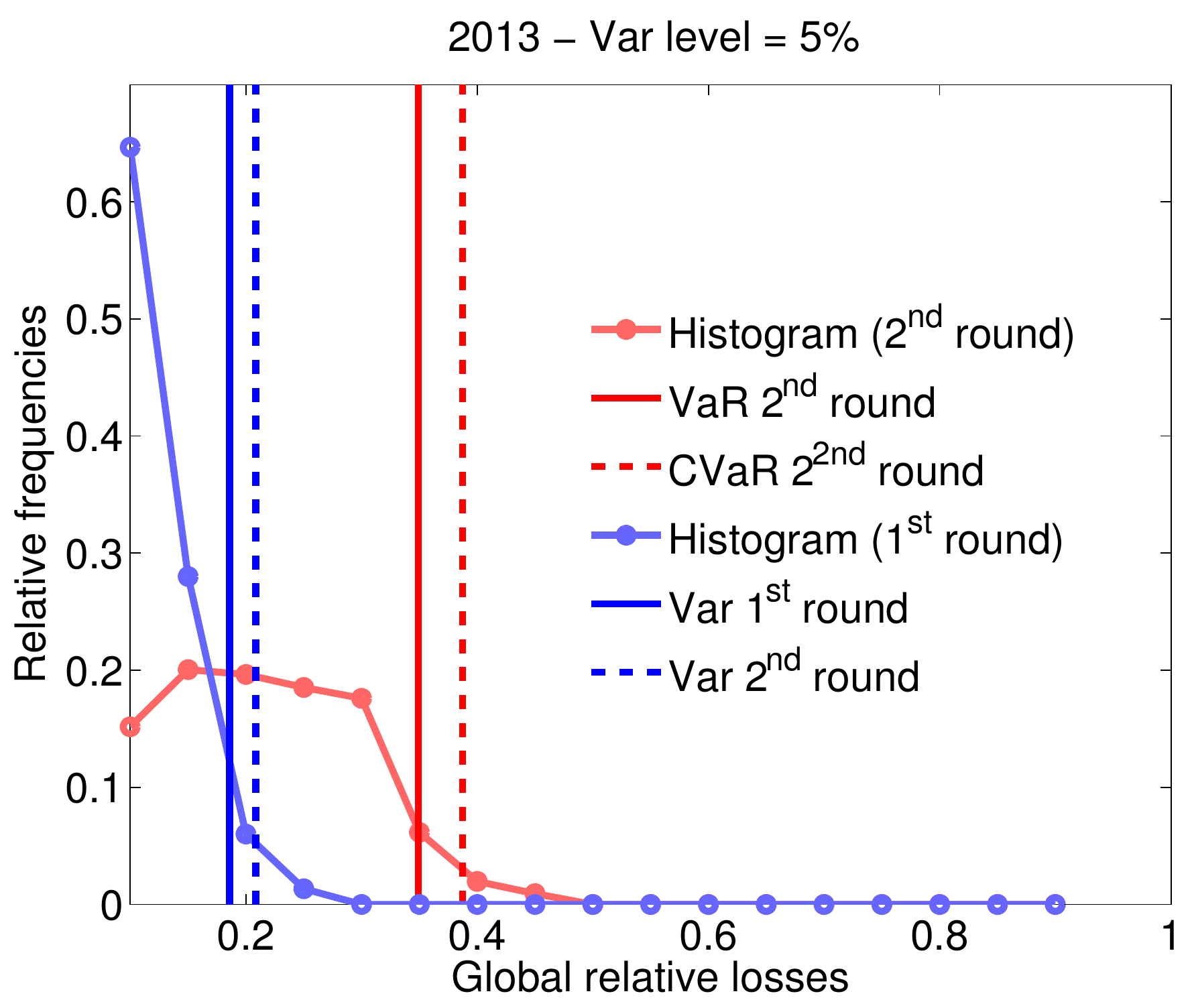}
\caption{Distribution of global relative losses (global vulnerability) in 2008 and 2013. Relative shocks on value of  external assets drawn from a Beta distribution with parameters $[4, 8]$ and truncated with a maximum of $0.015$.\label{fig:glob_losses_distr}}
\end{figure}

We evaluate a distribution of relative global equity losses by simulating $150$ different systemic shock levels drawn from a Beta distribution.\footnote{The parameter of the Beta distribution chosen are $a = 4$ and $b = 8$ respectively. The distribution has been then truncated in order to attain a maximum value of $0.015 = 1.5\%$ and a minimum of $0.001 = 0.1\%$.} Figure \ref{fig:glob_losses_distr} shows the distributions resulting by taking into account first only (blue lines) and second round (red lines) distress propagation effects for the years $2008$ and $2013$. Vertical lines indicate VaR values at 95\%, dashed lines are  CVaR at the same level (see \ref{ssec:loss_distr} for details). An extremely important consideration can be made from this figure: accounting for second order effects greatly increases the likelihood of having larger global equity losses, thus \textit{shifting} VaR values towards the right. In $2008$, a scenario where only first order distress is induced leads to a relatively low VaR level. This, instead, reaches a much higher value after the second round effect is added. A similar, though less extreme, pattern is found in $2013$. The observed \textit{VaR shift} phenomenon is another compelling piece of evidence stating that systemic risk measures ought to take into account network effects.

\subsubsection{Individual losses}
\begin{figure}
\includegraphics[width = 0.49\textwidth]{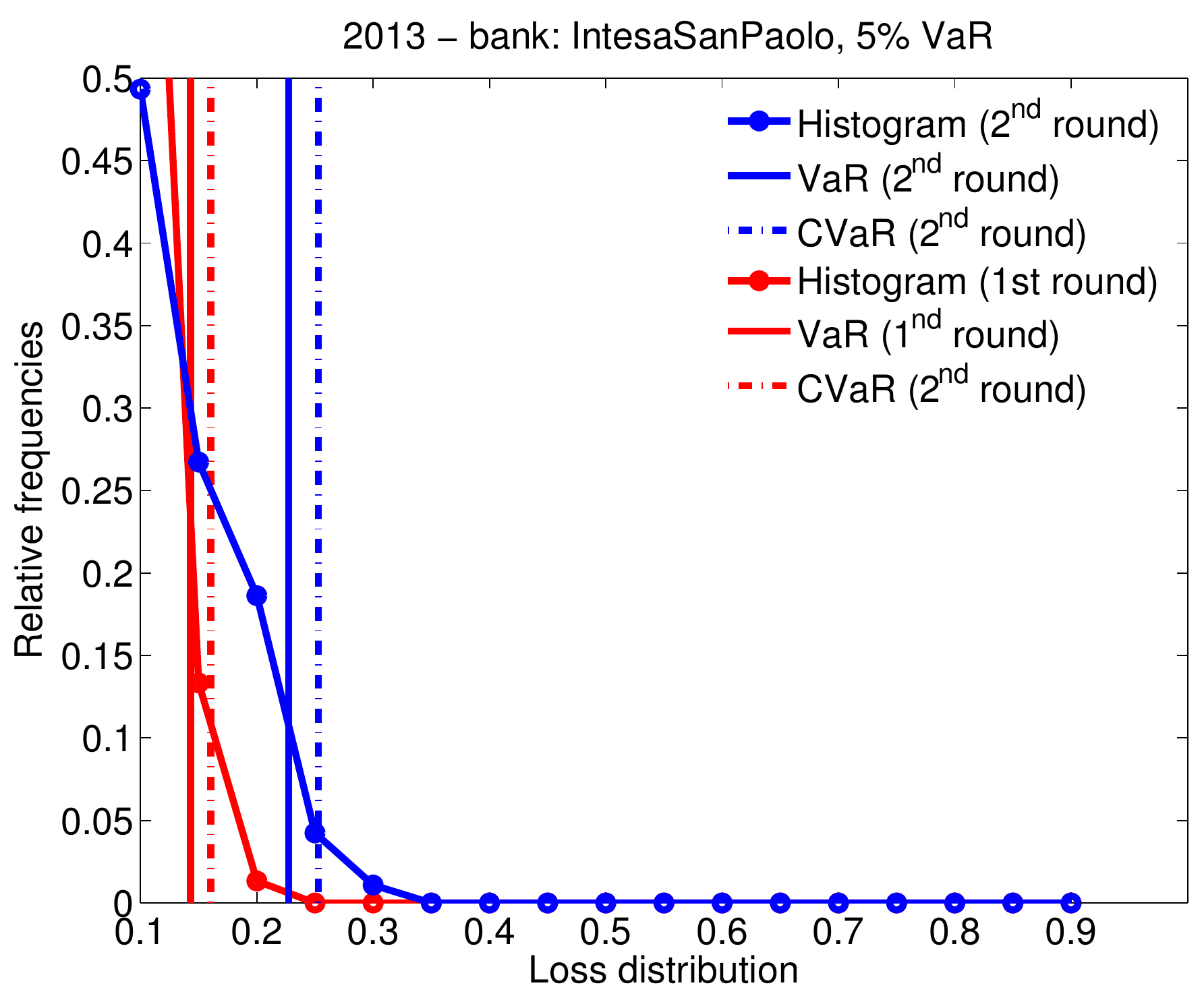}
\includegraphics[width = 0.49\textwidth]{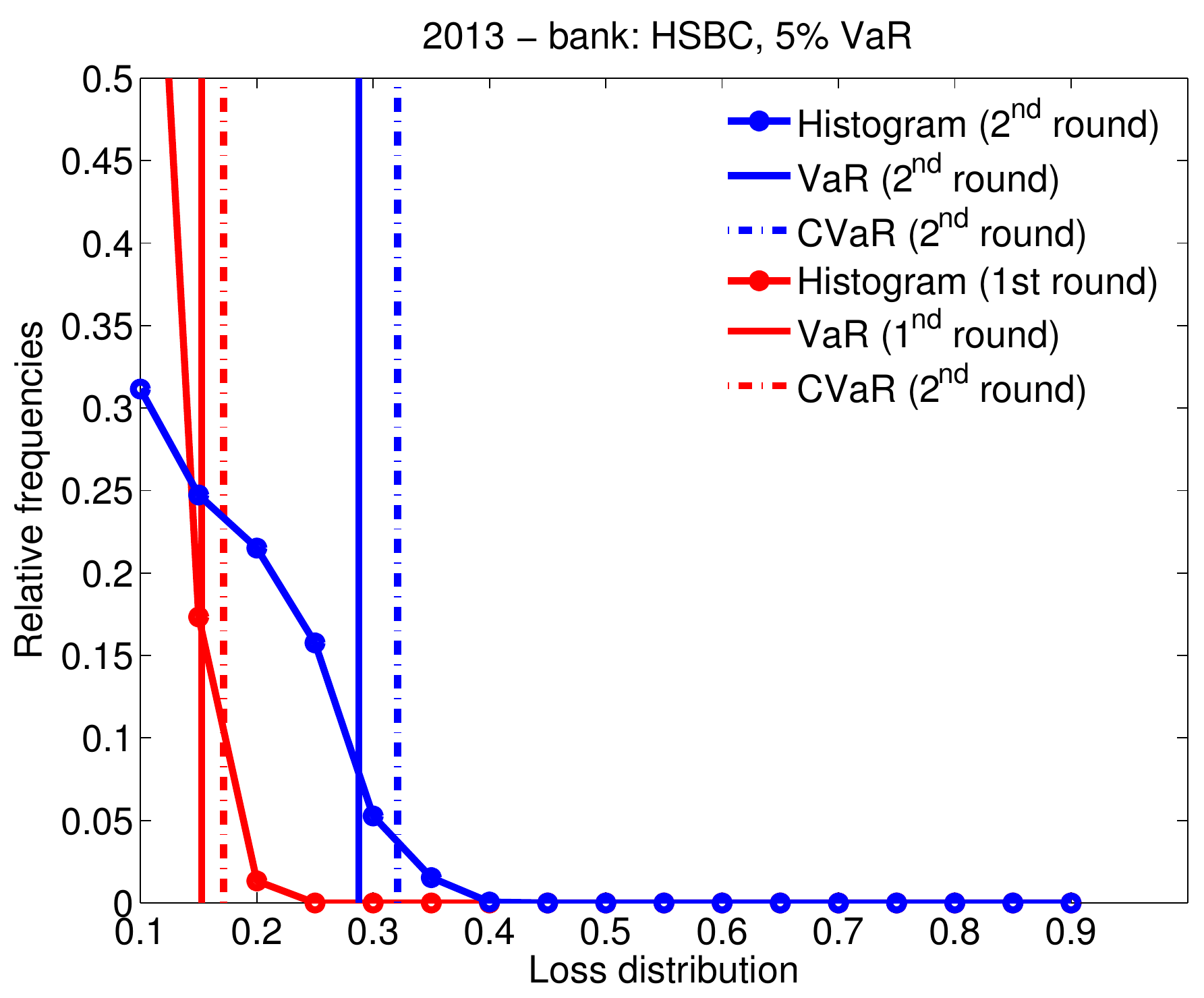}
\caption{Individual losses for two large banks. (Left) The chart reports the loss distribution for Intesa SanPaolo and (right) HSBC.\label{fig:individual_losses}}
\end{figure}

Figure \ref{fig:individual_losses} shows yet one of the outputs of the framework: the distribution of losses can be obtained for each individual bank. Here, we focus on two large institutions (by asset size): HSBC (which ranks first by asset size in $2013$) and Intesa SanPaolo (which ranks thirteenth in $2013$). Despite the difference in asset size, the original distance in the levels of VaR for the first round ($0.15$ vs $0.14$) become much more relevant when second round effects are considered ($0.28$ vs $0.22$). The example shows that significant differences in terms of standard risk measures are missed out if we neglect second-round effects.

\section{Discussion and concluding remarks\label{sec:discussion}}
\begin{figure}[h]
\includegraphics[width = 0.49\textwidth]{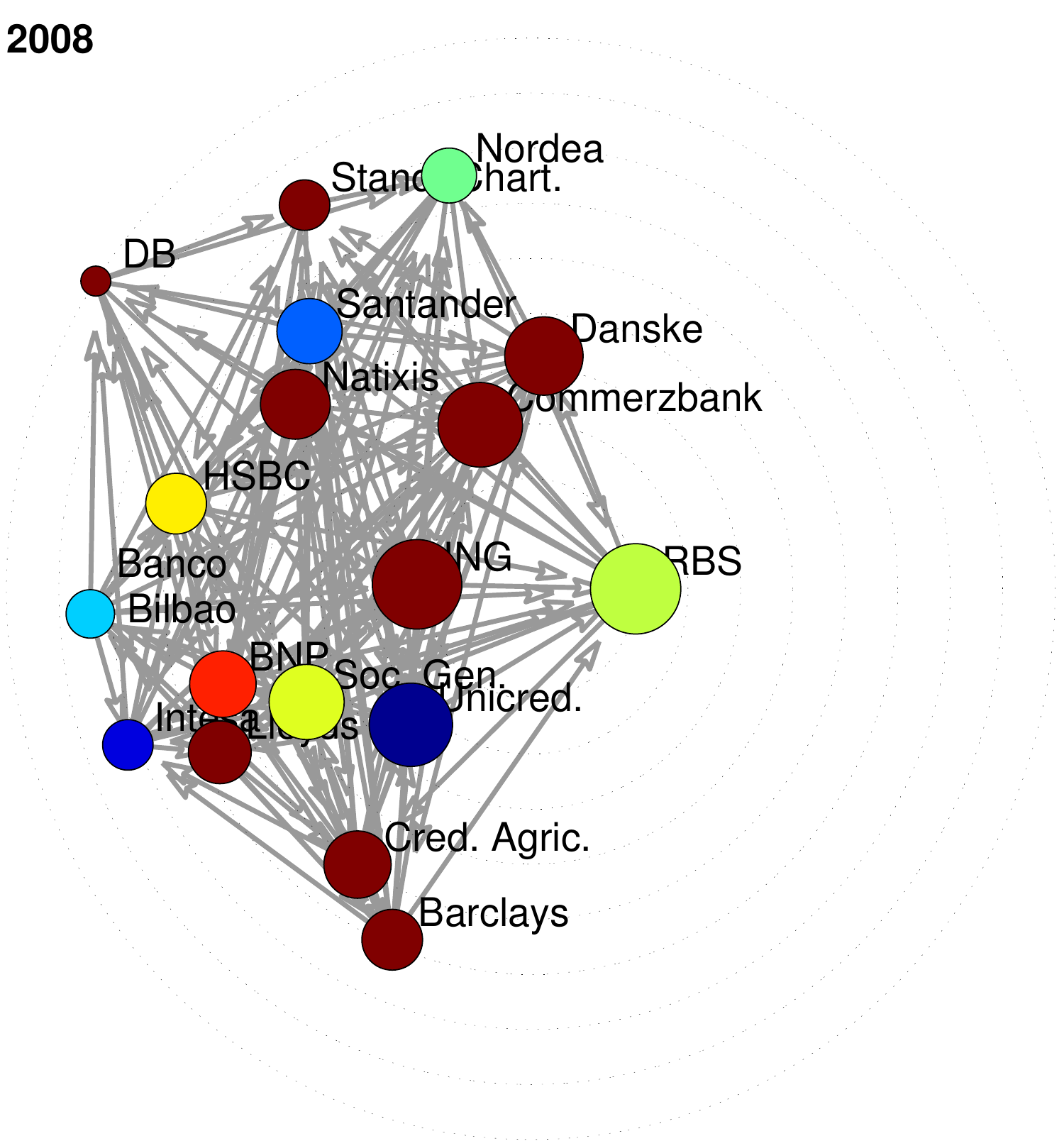}
\includegraphics[width = 0.49\textwidth]{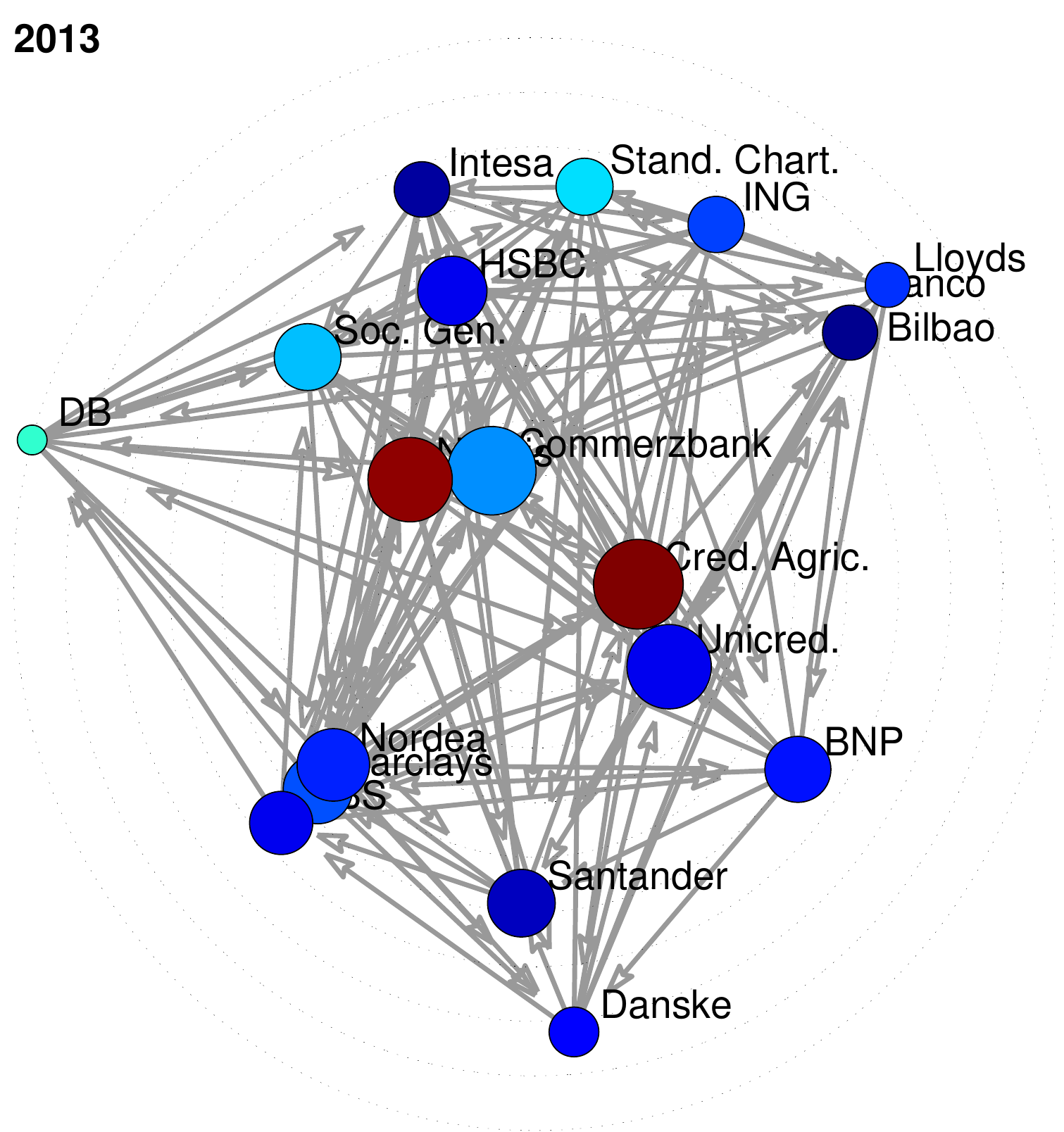}
\caption{Network visualization of the top 18 institutions by asset size in 2008 and 2013. Nodes are positioned on the concentric circles according to their Katz centrality.\label{fig:network_visualization}}
\end{figure}

The exercise carried out in Section \ref{sec:exercise} shows how the framework can be used to compute a variety of individual and global quantities that are relevant to systemic risk. The framework allows for a number of additional analyses which are not reported in detail in this paper for the sake of conciseness.  For instance, Figure \ref{fig:network_visualization} represents one of the outputs of the framework in terms of network \textit{visualization} and allows to compare the network position of individual institutions with other information. In this example, the interbank exposures among the top $18$ banks by total asset size in $2008$ (left) and $2013$ (right) are considered. The position of a bank in the chart is determined by its impact: the higher the impact, the more central the bank is located in the circle. The bubble size is proportional to total asset size of the bank, while the color encodes its vulnerability (on a scale from blue to red, red nodes are more vulnerable). 
It is worth mentioning the discussion on the determinants of the \textit{systemic importance of financial institutions}. In particular, one question is to what extent the asset size of an institution can be a good predictor of the impact of the bank on the system as a whole, and how much we should instead consider the position of the bank in the interbank network. Previous work have found that, although systemically important banks are typically among the large banks, banks with similar size can have very different impact on the system, in case of default \citep{diiasio2013capital}.  In line with those results, in our exercise, we find, loosely speaking, that asset size is \textit{not} a good predictor of impact (i.e. the Pearson correlation between asset size and individual impact, as measured by DebtRank, for the top $30$ institutions by total assets, each year is quite low, around $0.5$). 

To summarize, this paper presents a stress-test framework focused on the evaluation of network effects in systemic risk. We have illustrated how to carry out a stress-test exercise on a dataset of 183 European banks over the years $2008$-$2013$. The code underlying the framework has been developed in MATLAB and is available upon request to the authors. 

The notion of interconnectedness has already entered the  debate on ``Global Systemically Important Banks'' \citep[G-SIBs,][]{bis2011gsib}. However, this notion has been  meant so far in an aggregate sense, without fully recognizing that institutions with similar aggregated exposures can have very different levels of systemic impact and/or vulnerability to shocks. Indeed, a central notion in our framework is the one of \textit{leverage networks}, i.e. the set of leverage relations among banks' balance-sheets and among banks and assets. The effect of these relations is the key starting point to monitor systemic risk from a network perspective. Accordingly, our framework allows to track separately the magnitude of the so-called first, second and third round effects, a feature that is particularly important in the discussion of future stress-tests at national and international level. In this respect, in line with previous work on German interbank data \citep{fink2014price}, we find that the second-round effect is at least as large as the first-round effect. 

Notice that by adopting traditional analyses based on default-only mechanisms \citep{eisenberg2001systemic,rogers2013failure}, we would instead find very limited second round effects. There are two main reasons for this discrepancy. First, in a default-only framework the propagation of distress only occurs in the case of outright default. For example, in order to trigger any second round effect, at least one default in the first round is necessary: this  would in turn imply that an initial shock needs to be at least as large as the reciprocal of a bank's leverage (see Appendix \ref{sec:methods}). In order for the second round to be significant, there would need to be a large number of defaults in the first round. Indeed, banks are recommended to keep their single largest exposure well below their capital so that a necessary condition for second round losses is the default of at least two of their counterparties. However, in the practice, banks regularly re-evaluate their mark-to-market exposures to their counterparties in order to take into account the changes in their probability of default \citep[i.e. Credit Valuation Adjustment; see][]{bcbs2013cva}. Indeed, DebtRank captures the effects of this adjustment in a recursive way.  The second reason for the discrepancy arises from the fact that,  in \cite{eisenberg2001systemic}, while the recovery rate on interbank assets is determined endogenously, the recovery rate on the external assets is assumed to be one. Although this assumption has been relaxed in \citep{rogers2013failure},  very large shocks at the first round are still necessary to trigger any second round. In contrast, the dynamics of DebtRank assumes zero recovery rate on interbank assets, which is a realistic assumption in the short run. Future work will aim at bridging  these two  paradigmatic approaches.

In the framework, we further compute a series of systemic risk variables, along with their evolution over time, thus showing the dynamics of systemic risk in the financial system. In this respect, there is an added value in looking at quantities such as impact and vulnerability of financial institutions in combination, since systemic risk emerges when institutions that are systemically important become also vulnerable. While the results illustrated here have been obtained assuming the distress propagation mechanism of DebtRank \citep{battiston2012debtrank}, other mechanisms can also be used in the framework and compared. 


One of the obstacles in estimating network effects is the limitation in the availability of interbank exposures data. In order to address this issue, our framework allows to generate sets of interbank networks that satisfy the constraints on the total lending and borrowing of each bank. In this way, we can gain insights on the possible range of variation on systemic risk, due to differences in key network quantities. For example, one could tune the density parameter to assess whether this has an impact on the levels of systemic risk (see Appendix \ref{sec:network_reconstruction}), or use various interbank network formation models \citep[see, e.g.][]{halaj2015modelling}.

Overall, our aim is to enrich the set of existing tools by integrating the estimation of network effects with risk measures that are familiar to regulators and practitioners. The most used risk measures (such as Value at Risk and Expected Shortfall) look at the buffer that each individual bank needs to set aside in order to cover the \textit{direct} exposures to different types of shocks. In contrast, the \textit{indirect} exposures arising from the interconnected nature of the financial system are typically not considered in such measures. In this respect, our framework allows to estimate individual and aggregate banks' loss distributions conditional to both direct shocks and indirect shocks on other banks.

\appendix

\section{Methods}
\label{sec:methods}

In this methodological Appendix, we provide the technical details of the process underlying the stress-test framework. In order to bridge between capital requirements and the network structure, we build on the common notion of \textit{leverage} and define two \textit{leverage networks}, which reflect a more granular representation of banks' balance sheets.

\subsection{Balance-sheet dynamics}

In the framework, we consider a financial system composed of $n$ institutions (banks). Each institution $i$ in the system can invest in either $m$ external assets or in the funding of the other $n-1$ financial institutions. The focus of our analysis is on the dynamics of the balance sheets of each institution (at each time $t = 0, 1, 2, \ldots$) and, in particular, of their equity levels. The balance sheet is modelled as follows: $E_i(t)$ is the equity value of institution $i$ at time $t$, $A_i(t)$ is value of its total assets and $D_i$ its total liabilities. Consistently with much of the literature, we assume that assets are marked-to-market whereas liabilities are written at their face value. We can classify assets and liabilities into \textit{external} and \textit{interbank}. In particular, we consider the $n \times n$ interbank lending matrix, whose element $A_{ij}^b$ is the amount bank $i$ lends to bank $j$ in the interbank market and the $n \times m$ external assets matrix, whose element $A_{ik}^e$ is the amount invested by bank $i$ in the external asset $k$. The sum $A_i^b = \sum_{j = 1}^n A_{ij}^b$ is the total amount of interbank assets of bank $i$ and the sum $A_i^e = \sum_{k = 1}^m A_{ik}^e$ is the total amount of external assets of bank $i$. In this framework, we consider external liabilities as exogenous and do not specifically model them: to simplify the notation, these liabilities do not carry a time index. The balance sheet identity at each time $t = 0$ reads: $A_i(t) = D_i(t) + E_i(t)$, or, equivalently, $A_i^e (t) + A_i^b (t) = D_i^e + D_i^b(t) + E_i(t)$. We define the \textit{total leverage} of bank $i$ at time $t$ as the ratio between its total assets and its equity: $l_i(t) = A_i(t)/E_i(t)$, which can disaggregated into its additive subcomponents:

\begin{align}
l_i(t) & =  \frac{A_i(t)}{E_i(t)} =  \nonumber\\
& =  \frac{A_{i1}^b(t) + \ldots + A_{ij}^b(t)+ \ldots + A_{in}^b (t)+ A_{i1}^e (t)+ \ldots + A_{ik}^e(t) + \ldots +A_{im}^e(t)}{E_i(t)}  \nonumber \\
& = l_{i1}^b(t)+ \ldots + l_{ij}^b (t)+ \ldots + l_{in}^b(t) + l_{i1}^e(t) + \ldots + l_{ik}^e (t)+ \ldots + l_{im}^e(t)\label{eq:summands}
\end{align}

where the  element $l_{ij}^b(t) = A_{ij}^b/E_i(t)$ is the leverage of bank $i$ towards bank $j$ at time $t$ and the  element $l_{ik}^e(t) = A_{ik}^e/E_i(t)$ is the external leverage of bank $i$ with respect to the external asset $k$. By considering these two matrices as weighted adjacency matrices, we can then envision two \textit{leverage networks}: i) a mono-partite interbank leverage network and ii) a bipartite external leverage network. By summing along the columns of these matrices, we can obtain the \textit{total interbank leverage} $l_i^b(t) = \sum_{j} l_{ij}^b(t)$ (the interbank leverage \textit{out-strength}) and the \textit{total external leverage} $l_i^e = \sum_k  l_{ik}^e(t)$ (the external leverage \textit{out-strength}). These quantities are the key variables in our framework. In particular, we will show that interbank and external leverage produce compounded effects when the dynamic of losses for the second round is considered.

\subsection{The distress process}
As banks deplete capital in order to face losses in both interbank and external assets, in the stress-test framework we are mainly concerned with the dynamics of the relative loss in equity for each institution, with respect to a baseline level at $t = 0$. This dynamics is captured by the following process:

\begin{equation}
\label{eq:individual_equity_loss}
h_i(t) =\min\left\{ 1,  \;\; \frac{E_i(0) - E_i(t)}{E_i(0)} \right\}, \;\; t = 0, 1, 2, \ldots.
\end{equation}
which represents the individual cumulative relative equity loss in time. We assume that either no replenishment of capital or positive cash flow are possible, therefore $E_i(t) \leq E_i(t-1)$, $\forall t$. In this way, the relative equity loss is a non-decreasing function of time. Further, $h_{i}(t)  \in [0, 1]$  $\forall t$. A bank defaults ( i.e. the bank reaches the maximum distress possible)  if $h_i(t) = 1$. When $h_i(t) = 0$ the bank is undistressed. All values of $h_i(t)$ between $0$ and $1$ imply that the bank is under distress. Similarly, we can compute the global cumulative relative equity loss at each time $t$ as the weighted average of each individual level of distress:

\begin{equation}
\label{eq:global_equity_loss}
H(t) = \sum_i  w_i \; h_i(t)
\end{equation}

where the weights are given by $w_i = E_i(0)/\sum_j E_j (0)$, i.e. the fraction of equity of each bank at the baseline level ($t = 0$). Notice  that $h_i(t)$ is a pure number and so is $H(t)$.  The monetary value (e.g. in Euros or Dollars) of the loss can be obtained by $h_i(t) \times E_i(0)$ (individual loss) and $H_i(t) \times \sum_i E_i(0)$ (global loss). 

Using the terminology introduced in the main text, Equations \ref{eq:individual_equity_loss} and \ref{eq:global_equity_loss} allow to measure the \textit{individual} and \textit{global} vulnerability respectively.  The entire distress process featured in the framework can be outlined in the following steps.

\subsubsection{First round: shock on external assets}

Let $p_k(0)$ be the value of one unit of the external asset $k$. At time $t = 1$, a (negative) shock $r_k(1) = \frac{p_k(0) - p_k(1)}{p_k(0)}$ on the value of asset $k$ reduces the value of the investment in external assets of bank $i$ by the amount: $\sum_k r_k(1) A_{ik}  = \sum_k r_k(1) \; l_{ik} E_{i} = E_i \sum_k r_k(1) \; l_{ik}$. Banks record a loss on their asset side that, provided the hypothesis that assets are mark-to-market and liabilities are at face value, the loss needs to be compensated by a corresponding reduction in equity:

\begin{equation}
\nonumber
A_{ik}^e(0) - A_{ik}^e(1) = \sum_k r_{k}(1) \; A_{ik}^e(0)  = E_i(0) - E_i(1)
\end{equation}

The individual and global relative equity loss at time $t = 1$  can be obtained as follows:\footnote{We assume that the write off on the value of external assets is entirely absorbed by the equity; the derivation is straightforward: $$h_i(1) = \min \left\{1, \; \frac{E_i(0) - E_i(1)}{E_i(0)}\right\} = \min \left\{1, \; \frac{\sum_k A_{ik}^e(0) r_k(1)}{E_i(0)}\right\} = \min \left\{1, \; \sum_k \left( l_{ik}^e  \times r_k(1)\right)\right\}.$$}
\begin{equation}
\nonumber
h_i(1) = \min \left\{1, \; \; \sum_k l_{ik} r_k(1) \right\} \text{ and } \; H(1) = \sum_{i = 1}^n  w_i \; h_i(1),
\end{equation}

which shows how the initial shock on each asset $k$ is \textit{multiplicatively} amplified by the external leverage on that specific asset. This leads to a straightforward interpretation of the leverage ratio. Indeed it is immediate to prove that the reciprocal of the leverage ratio corresponds to the minimum shock $r_i^{\text{min}}$ that leads bank $i$ to default (this applies to all  summands $l_{ik}^e$ e $l_{ij}^b$ in Equation \ref{eq:summands}). Since the single largest exposure is typically smaller than the equity, it is likely that defaults and large losses  originate by different combinations of shocks affecting the different external assets. 
In the absence of detailed data on the exposure to different classes of external assets, we assume a \textit{common} negative shock $r(1)$ on the value of all external assets. This assumption can be interpreted in two alternative ways. First, we can envision a common small shock to all asset classes, as in times of general market distress. The second way is that of a large shock to specific asset classes held by all banks (e.g. sovereign on a class of countries, housing shocks, etc.).

We can therefore drop the index $k$ in the summation and write: $h_i(1) = \min\{1, l_i^er(1)\}$. At this point, the initial loss reverberates throughout the interbank network.

\subsubsection{Second round: reverberation on the interbank network}
The DebtRank algorithm \citep{battiston2012debtrank} extends the dynamics of default contagion into a more general distress propagation not necessarily entailing a default event. In other words, shocks on the asset side of the balance sheet of bank $i$ transmit along the network even when such shocks are not large enough to trigger the default of $i$. This is motivated by the fact that, as $i$'s equity decreases, so does its \textit{distance to default} \citep{kmv2003distance} and, consistently with the approach of \cite{merton1974pricing} the bank will be less likely to repay its obligations in case of further distress, therefore implying that the market value of $i$'s obligations will decrease as well. Consequently, the distress propagates onto its counterparties along the network. If we denote the market value of the obligation with $V_t(A_{ij})$,\footnote{From a balance sheet perspective, $A_{ij}$ is the element standing on the liability side of $j$ (i.e. the face value established at time $0$), whereas $V_t (A_{ij})$ is the value (mark-to-market) at time $t$ written on the asset side of i.} then above argument implies that the distress  $j$  propagates onto its lender $i$ can be expressed, in general terms, as the relative loss with respect to the original face value $\frac{A_{ij} - V_t(A_{ij})}{A_{ij}} = f(h_j(t-1))$.  By summing over all obligors, the relative equity loss of each bank $i$ at time $t = 2, 3, \ldots$ is described by:

\begin{equation}
\label{eq:main_debt_rank}
h_i(t) = \min \left\{1, \sum_{j \in S_A(t) } l_{ij} f(h_j(t-1))\right\}
\end{equation}

where $S_A(t)$ is the set of \textit{active} nodes, i.e. nodes that transmit distress at time $t$. The choice of the set of active nodes at time $t$, $S_A(t)$, is a peculiarity of DebtRank. In fact, Equation \ref{eq:main_debt_rank} is of a recursive nature and therefore needs to be computed at each time $t$ by considering the nodes that were in distress at the previous time. Since the leverage network can present cycles, the distress may propagate via a particular link more than once. Although this fact does not represent a problem in mathematical terms, its economic interpretation is indeed more problematic. In order to overcome this problem, DebtRank excludes more than one reverberation. From a network perspective, by choosing the set $S_A(t)$ we exclude walks that count a specific link more than once. The process ends at a certain time $T$, when nodes are no longer active.

\paragraph{The functional form of $f(\cdot)$.} The choice of the function $f(\cdot)$ deserves further discussion. In fact, a correct estimation of its form would require an empirical framework which should take into account the probability of default of $j$ and the recovery rate of the assets held by $i$. However, the minimum requirement that $f(\cdot)$ needs to satisfy is that of being a non-decreasing relation between $h_i$ and the losses in the value of its obligations. More specifically, we can hypothesize that small values of $h_i$ may have little to no effect on the market value of $i$'s obligations, whereas extremely large losses would settle the value of $i$'s obligations almost close to zero: the relationship is therefore necessarily non-linear and $f(\cdot)$ is likely to be a sigmoid-type of function. In view of this, although further work will deal with the analysis of more refined functional forms, we hereby present two main forms, referring to the following two specific  dynamics of distress:
\begin{description}
\item[Default contagion.] In this case, in line with a specific stream of literature, \citep{eisenberg2001systemic}, only the event of default triggers a contagion. The function $f(\cdot)$ is therefore chosen as the  indicator function over the case of default $f(h_i(t)) = \chi_{\{h_i(t) = 1\}}$. 


\item[DebtRank.] The characteristics of $f(\cdot)$ imply the existence of an intermediate level where $f(\cdot)$ can be approximated by a linear function. By choosing the identity function $f(h_i(t)) = h_i(t)$,  we  obtain to the original DebtRank formulation \citep{battiston2012debtrank}. This functional form will be the one we use the most in the framework and the exercise. 
\end{description}

For the sake of clarity, in the remainder of this Section, we consider only the latter functional form. However, in the framework, stress tests can be easily carried out for both cases. 

\paragraph{Vulnerability.} We are now ready to compute the vulnerability (both individual and global) and the \textit{impact} (at the individual level). The individual vulnerability $h_i(t)$ can be easily computed by setting $f(h_j(t)) = h_j(t)$ in Equation \ref{eq:main_debt_rank}. The global vulnerability is then given by $H(t) = \sum_{i} h_i(t) w_i$. Even though the framework can take as input \textit{any} type of shocks, we focus briefly on the case in which the external assets of all banks are shocked: in this case all banks transmit distress at time $t = 1$ and, given the choice of the set $S_A(1)$, the process indeed ends at time $T = 2$. We can hence derive a closed-form solution for  the individual vulnerability after the second round:
\begin{equation}
\label{eq:first_order_approx}
h_i(2) = 	\min \left\{1, l_i^e r(1) + \sum_j l_{ij}^b l_j^e r(1)\right\},
\end{equation}
which elucidates the \textit{compounding} effect of external and interbank leverage. If the shock $r(0)$ is small enough not to induce any default, then \ref{eq:first_order_approx} can be rewritten as:
\begin{equation}
h_i(2)   = 	 l_i^e r(1) + \sum_j l_{ij}^b l_j^e r(1)  = r(1) \left(l_i^e + \sum_j l_{ij}^b l_j^e\right) \nonumber
\end{equation}

\paragraph{Impact.} DebtRank, in its original formulation \citep{battiston2012debtrank}, entails a stress test by assuming the default of each bank individually and computing the global relative equity loss \textit{induced} by such default. This is indeed what we define as the \textit{impact} of an institution onto the system as a whole. Formally, this can be written as:
\begin{equation}
\label{eq:impact}
DR_k = \sum_i h_i(T) E_i(0).
\end{equation}

\paragraph{Network effects: a first order approximation of vulnerability}
Equation \ref{eq:main_debt_rank} clearly shows the main feature of the distress dynamics captured by DebtRank: the interplay between the network of leverage and the distress imported from neighbors in this network. Further, Equation \ref{eq:first_order_approx} clarifies the multiplicative role of leverage in determining the distress at the end of the second round. We now develop a first-order approximation of Equation \ref{eq:first_order_approx}, which will serve the purpose of further clarifying  the \textit{compounding} effects of external and interbank leverage in determining distress.  For the sake of simplicity, we assume no default, which allows us to remove the ``$\min$'' operator. This is a reasonable assumption in case of a relatively small shock on external assets. We approximate the external leverage of the obligors of bank $i$ by taking the weighted average (with weights $w_i$) of their external leverages, which we denote by $l^e$. As $\sum_j l_{ij}^b = l_i^b$, we write $h_i(2) \approx l_i^e r + l_i^b \; l^e \; r$.  By denoting with $l^b$ the weighted average of $l_{i}^b$, we can approximate the global equity loss at the end of the second round $H(2)$ as:
\begin{equation}
\label{eq:debt_rank_approx}
H(2) \approx l^e r +l^b \; l^e \; r\,
\end{equation}
which allows to see how the second-round effects alone can be obtained as the product of the weighted average interbank leverage and weighted average external leverage. Typically, stress tests emphasize the effects of the first-round: as we observe, this may potentially bring to a severe underestimation of systemic risk.

\subsection{Third round and fire sales}

After the second round, banks have experienced a certain level of equity loss that has completely reshaped the initial configuration of the balance sheets at time $t = 0$. Banks are now attempting to restore, at least partially, this initial configuration. In particular, we assume \citep{tasca2013procyclicality} that each bank $i$ will try to move to the original leverage level $l_i(0)$. This implies that banks will try to sell external assets in order to obtain enough cash to repay their obligations and therefore reduce the size of their balance sheet. Because of the vast quantity of external assets sold by the banking system in aggregate, the impact on the prices of external assets is also relevant, which will reduce accordingly. Banks therefore will experience further losses due to fire sales and we label such losses as \textit{third round} effects. Here, we provide a minimal model for the scenario described above.

Consider the leverage dynamics at  $t =1, 2, \ldots, T, T + 1, T+2$. The leverage at $t$ is 

\begin{equation}
l_i(t) = l_i^e(t) + l_i^b(t) = \frac{A_i^e(t) + A_i^b(t)}{E(t)} 
\end{equation}

We assume that, at $t = 0$, each bank had a quantity of external assets $Q_i$ and, without loss of generality, that the initial price of the asset is unitary ($p(0) = 1$). Hence, the asset values at $t = 0$ can be written as $A_i(0) = Q_i(0) = l_i(0) E_i(0)$. The asset price after the first round is therefore simply $p(1) = p(T) = (1-r)$. Recalling that the first round affects only the external asset and that the second round affects only interbank assets, the leverage of each bank $i$ immediately after the second round can be written as:
\begin{align}
l_i(T) & = \frac{(1-r) Q_i + A^b_i(0) - (h_i(2) - h_i(1))}{(1 - h_i(2)) E_i(0)}= \nonumber\\
& = \frac{(1-r) l^e_i E_i(0) + l_i^b E_i(0) - (h_i(2) - h_i(1))E_i(0)}{(1-h(2)) E_i(0)}= \nonumber\\
& = \frac{(1-r) l^e_i + l_i^b - (h_i(2) - h_i(1))}{1 - h_i(2)}= \nonumber\\
& =  \frac{(1-r) l^e_i + l_i^b - h_i(2) + l_i^e r}{1 - h_i(2)} \label{eq:newleveragelevels}
\end{align}

where, for ease of notation $ l_i^e =  l_i^e(0)$ and $ l_i^b =  l_i^b(0)$. First, we need to prove that the new leverage levels are higher with respect to the initial conditions. It is easy to prove that $l_i(T) > l_i(0)$ (as long as $i$ has not defaulted):
\begin{equation}
(1-h_i(2))(l^e_i + l_i^b - h_i(2) + l^e_i r) > (1-h_i(2)) (l_i^e + l_i^b) \Longleftrightarrow (1-h_i(2)) l_i < l_i - h_i(2) \nonumber
\end{equation}
where $l_i = l_i(0)$. The above inequality leads to the condition $h_i(2)(l_i -1) > 0$, which is always verified in our setting.

At $t =  T + 1$, banks attempt to restore the target leverage $l_i^* = l_i(0) = l_i^e + l_i^b$, by selling a fraction $s_i \in [0, 1]$ of their external assets at the price $(1-r)$ and replenish their equity of an amount $Q_i (1-r) s$. Therefore, we modify Equation  \ref{eq:newleveragelevels} as follows:

\begin{equation}
l_i^e + l_i^b = \frac{(1-s_i)(1-r)l_i^e + l_i^b - h_i(2) + l^e_i r}{(1-h_i(2)) + s_i(1-r)l^e_i}\label{eq:newnewleverage}
\end{equation}

After some passages, we obtain the value for $s_i$:

\begin{equation}
s_i = \frac{h_i(2)}{(1-r) l_i^e} \;\frac{ l_i - 1} {l_i + 1} \in (0, 1) \nonumber
\end{equation}
which satisfies Equation \ref{eq:newnewleverage}. The relative amount of assets sold is given by  $\rho = \frac{\sum_i s_i A_i^e}{\sum_i A_i^e}$. We further assume that the simultaneous selling of external assets in the market produces a further linear impact on the price. Given the impact of fire sales, the new price is further reduced as follows: 

\begin{equation}
p(T + 2) = (1-r) (1 - \rho  \eta)  \label{eq:pricedecline}
\end{equation}

and the relative change in price is therefore proportional to the relative change in quantity of sold assets through a constant $\eta \in [0, 1]$. Finally, by computing the additional loss given by the decline in price following Equation \ref{eq:pricedecline}, we obtain the final individual relative equity loss at $t = T + 2$:

\begin{align}
h_i(T+2) & = \min\left\{1, h_i(T) + l_i^e(1-r)(1-s_i) \rho \eta\right\} = \nonumber\\
& = \min\left\{1,  l_i^e r + \sum_j l_{ij}^b l_j^e r  + l_i^e(1-r)(1-s_i) \rho \eta \right\}\nonumber\\ 
\end{align}

and the global equity loss at the third round (assuming no defaults):

\begin{align}
H(T+2) & = H(2) +  (1-r)  \rho \eta \sum_i \left(w_i  l_i^e(1-s_i) \right) \nonumber \\
& = \sum_i w_i \left(l_i^e r + \sum_j l_{ij}^b l_j^e r\right)+  (1-r)  \rho \eta \sum_i \left(w_i  \; l_i^e(1-s_i) \right)
\nonumber
\end{align}

\subsection{Loss distribution\label{ssec:loss_distr}}
The distress process allows to capture, at each time $t$, the relative equity loss for both the individual institution and the system as a whole. This implies the possibility to compute, at each time $t$, a (continuous) \textit{relative equity loss distribution} conditional to a certain shock. The  equity loss distribution can be characterized, for example, by two typical risk measures: Value at Risk (VaR) and Conditional Value at Risk (CVaR) (also known as Expected Shortfall, ES). Since $h_i(t)$ and $H(t)$ are nonnegative variables $\in [0, 1]$ $\forall i, t$, the individual Value at Risk for bank $i$ at time $t$ at level $\alpha$ is defined as the $1 - \alpha$ quantile  \citep{mcneil2010quantitative, follmer2011stochastic}:
\begin{equation}
\label{eq:var}
VaR^\alpha_{i}(t) = \inf \{x \in [0, 1]: P(h_i(t) \leq x) 	\geq (1 - \alpha)\}
\end{equation}
and the Conditional Value at Risk for  bank $i$ at time $t$ at level $\alpha$ is defined as the expected value of the losses exceeding the VaR, as:
\begin{equation}
\label{eq:cvar}
CVaR_i^{\alpha}(t) =E \left[h_i(t) | h_i(t) \geq VaR^\alpha_{i}(t)\right]
\end{equation}
Considering the system as a whole, we can likewise analyze the global relative equity losses $H(t)$ at each time $t$, therefore obtaining a \textit{global VaR}:
\begin{equation}
\label{eq:globvar}
VaR^\alpha_{\text{glob}}(t) = \inf \{x \in [0, 1]: P(H(t) \leq x) \geq (1 - \alpha)\},
\end{equation}
and the \textit{global CVaR}:
\begin{equation}
\label{eq:globcvar}
CVaR^\alpha_{\text{glob}}(t) =E \left[H(t) | H(t) \geq VaR^\alpha_{\text{glob}}(t) \right].
\end{equation}

\section{Data collection and processing\label{sec:data_collection}}
Detailed public data on banks' balance sheets are unavailable, therefore we resorted to a dataset that provides a reasonable level of breakdown, the Bureau Van Dijk Bankscope database (URL: \verb+bankscope.bvdinfo.com+). We focus on a subset of $183$ banks headquartered in the European Union that are also quoted on a stock market for the years from 2008 to 2013. The main criterion for the selection was that of having detailed coverage (on a yearly basis) for total assets, equity, interbank lending or borrowing.\footnote{In details: we recorded the fields 1) ``Equity'', 2) ``Total Assets'', 3) ``Total Liabilities and Equity'',  4) ``Loans and Advances to Banks'', 5) ``Deposits from other banks'' from the Universal Banking Model (UBM) of Bankscope. See \url{https://www.bvdinfo.com/getattachment/a5a81707-c96d-4525-9142-7c7e607abf56/Bankscope} and \url{http://www.bvd.co.uk/bankscope/bankscope.pdf}} Future work will deal with data at higher frequency (quarterly, monthly, \ldots). Our interbank asset and liability data include amounts due under repurchase agreements (which are economically analogous to a secured loan) thereby prompting large contagion effects.  We perfomed a series of consistency checks. In the case of missing interbank lending data for a bank for less than three years, we proceed with an estimation via linear interpolation of the data available for the other years  (a comparison with the available data  gives errors lower than 20\%). Since, in general, the correlation between interbank lending and borrowing for all banks and years is about 70\% (with some significant differences), this implies the presence of net lenders and net borrowers. In view of this, when data on either interbank lending or borrowing are not available for more than three years, we simply set them equal.

\section{Network reconstruction\label{sec:network_reconstruction}}

Data on total interbank lending and borrowing are often publicly available, while the detailed bilateral exposures are typically confidential. However, in this Section, we outline the estimation procedure adopted in the framework. At each point in time, we create a sample of $100$ networks via the ``fitness model'', which is a technique that has recently been used to reconstruct  financial networks starting from aggregate exposures \citep{demasi2006fitness,musmeci2012bootstrapping,montagna2014contagion}. The procedure can be outlined as follows:

\textbf{1. Total exposure re-balancing.}
Since we are considering a subset of the entire interbank market, we observe an inconsistency: the total interbank assets $A = \sum_i A_i$ are systematically smaller than the total interbank liabilities $L = \sum_i L_i$ for each year (EU banks are net borrowers from the rest of the world). To adopt a conservative scenario, we assume that the total lending volume in the network is the minimum between the two ($A$ in the exercise). Let $A_i/A$ and $L_i/\sum_j L_j$ be respectively the lending and borrowing propensity of $i$.

\textbf{2. Exposure link assignment.} The fitness model, when applied to interbank networks \citep{demasi2006fitness} attributes to each bank a so-called fitness level $x_i$ (typically a proxy of its size in the interbank network). We can estimate the probability that an exposure between $i$ and $j$ exists via the following formula, $p_{ij} = \frac{z x_ix_j}{1+z x_i x_j}$ ($z$ is a  free parameter). Notice that $p_{ij} = p_{ji}$. Consistently with a recent stream of literature \citep{musmeci2012bootstrapping,montagna2014contagion}, for each bank we take as fitness $x_i$ the average between its total lending and  borrowing propensity, implying that,  the greater this value, the higher will be the number of counterparties (the \textit{degree} of a node). Considering  empirical evidence on the density of different interbank networks \citep{intveld2014finding}, we assume on average a density of $5\%$ (i.e. about $1670$  over the $n(n-1)$ possible links).\footnote{We have carried out a sensitivity analysis to assess the role of a specific choice of the density level. Increasing density to $10\%$ does not influence the overall results of the exercise. For example, values for the global vulnerability at the second round differs only at the third decimal digit.} Since it can be proved that  the total number of links is equal to the expected value of $\frac{1}{2} \sum_{i}\sum_ {j \neq i} \frac{zx_i\, x_j}{1+z\, x_i \,x_j}$, we can determine the parameter $z$ and compute the matrix of link probabilities $p_{ij}$. We now generate $100$ network realizations. For each of these realizations, we assign a link to the pair of banks $(i,j)$ with probability $p_{ij}$. The link direction (which determines whether  $i$ or $j$ is the lender or the borrower) is chosen at random with probability $0.5$. 

\textbf{3. Exposure volume allocation} Last, we need to assign weights to the edges (the volumes of each exposure). We impose the fundamental constraint that the sum of the exposures of each bank (out-strength) equals its total interbank asset $A_i$. To achieve this, we implement an iterative proportional fitting algorithm on the interbank exposure matrix $a_{ij}$. We wish to estimate the matrix $\pi_{ij} = A_{ij}/A$, which is the relative value of each exposure with respect to the total interbank volume. We begin the estimation $\hat{\pi}_{ij}$ of $\pi_{ij}$, at each iteration: (1) $\hat{\pi}^{\prime}_{ij} = \frac{\hat{\pi}_{ij}}{\sum_j \hat{\pi}_{ij}} \frac{A_i}{A}$, i.e. $\hat{\pi}_{ij}$ is divided by its relative lending propensity and multiplied by the total relative assets of $i$,; (2)  $\hat{\pi}^{\prime\prime}_{ij} = \frac{\hat{\pi}^{\prime}_{ij}}{\sum_i{\hat{\pi}^{\prime}_{ij}}} \frac{L_i}{L}$ $\hat{\pi}^{\prime}_{ij}$. We repeated the two steps until $\sum_j \hat{\pi}_{ij} - A_i/A$ and $\sum_j \hat{\pi}_{ji} - L_i/L$ are below $1\%$. Last, the exposure network can be estimated by $\pi_{ij} \times A$.

\end{document}